\documentclass[preprint,showpacs,preprintnumbers,amsmath,amssymb]{revtex4-2}
\usepackage{graphicx}
\usepackage{bm}
\usepackage{color}
\usepackage{xcolor}
\usepackage[colorlinks=true, allcolors=blue]{hyperref}
\usepackage{amssymb}
\usepackage{epsfig}
\usepackage{array,multirow}
\usepackage{float}
\newcolumntype{L}{>{\raggedright\arraybackslash}p{2.5cm}}

\usepackage{array,multirow}
\DeclareGraphicsExtensions{.png,.pdf}

\begin{document}

\title{Distribution of the entanglement entropies of non-ergodic quantum states}

\author{Devanshu Shekhar and Pragya Shukla$^*$}
\affiliation{ Department of Physics, Indian Institute of Technology, Kharagpur-721302, West Bengal, India }
\date{\today}

\widetext

\begin{abstract}

Beginning from an ensemble of pure bipartite states typically characterized by arbitrary, non-maximum entanglement, we seek the route to achieve an ensemble with  maximum entanglement  for a typical state. Our approach is based on the consideration of a pure quantum state, with its components non-identical but independent Gaussian distributed in a bipartite product basis. A change of ensemble parameters leads to a variation of the entanglement entropy distribution from its initial to maximum state. We show that the variation can be described by a common mathematical formulation for a wide range of initial ensembles, with a function of all ensemble  parameter governing the variation. The information provides an alternative approach to quantum purification and error correction for communication through noisy quantum channels.

\end{abstract}

\def\stackalignment{l}
\maketitle

{*Corresponding author, E-Mail: shukla@phy.iitkgp.ac.in}

\section{Introduction} \label{secIntro}

Quantum entanglement is a unique phenomenon, relevant not only for fundamental considerations but also for its increasing viability as a resource for various quantum information processes \cite{Dur,bdsw}.
A maximally entangled state is an ideal requirement for a quantum information process. While  entangled states can be produced under controlled experiments, unavoidable noise in such control operations, thermal fluctuations  as well as interactions with an uncontrollable environment leave the desired state with a non–unit fidelity. The search for routes to protect the entangled states and the information carried by them over noisy quantum  channels has led to various tools e.g. quantum error correction \cite{bdsw}, entanglement purification \cite{Dur} etc. Specific limitations associated with previous tools however motivate search for the new ones. The primary objective of present study is to suggest a new  tool  which results in an ensemble of maximum entangled states beginning from an ensemble of separable or partially entangled states.

Almost all protocols for maintenance of the high fidelity entanglement for  practical applications require consideration of an ensemble of states. For example, for long-range quantum communication, a route to protect  quantum information is  by its encoding, through entanglement, in a concatenated quantum error correcting code.  The transmission of the encoded quantum state through a noisy channel is followed by an error correction step that involves decoding and measurements. The noisy  passage however  randomizes the encoded state, leaving it best described by an ensemble.  For any efficient decoding, it is relevant to know a priori the entanglement distribution  of the ensemble.  
An alternative quantum communication route  is based on entanglement purification or distillation:  it manipulates an ensemble of noisy, non-maximally entangled states so as to leave  behind only a fewer number of copies with a reduced amount of noise. The entanglement of the total ensemble is concentrated or distilled in a few copies thereby containing a larger amount of entanglement and have higher fidelity.


In real applications, noise is usually unavoidable. It is thus natural to query whether it is possible to harness the noise to favour the quantum information processes? More clearly, is it possible to start from a pure state with arbitrary entanglement and apply noisy operations to convert it into an ensemble of fully entangled states?   An additional issue arises from a wide range of system conditions e.g. from external noise (changing environment) which can lead to ensembles with different types and strength of randomness. It will therefore be useful to have a protocol that can control  the dependence  collectively i.e through a single function of all system parameters. The present study pursues the above queries by  theoretically analysing the entanglement distribution for a wide range of ensembles of quantum states and seek a common mathematical formulation for its dynamics if feasible.  Our results are encouraging: we find that such a formulation indeed exists for pure states with Gaussian distributed components in a bipartite basis. The formulation in turn helps to identify the path the protocol can opt for to achieve maximum entangled ensemble  starting from a non-maximum entangled ensemble.

The standard representation of a quantum state is based on  its components describing the overlap of the state with basis vectors in a  physically motivated basis.
It is  however  technically non-trivial, almost impossible, usually, to determine the components exactly if the state is that of a many-body system \cite{wigner1967random}. A lack of exact information about the Hamiltonian matrix elements e.g. due to disorder or complicated interactions usually manifests through a  randomization of the eigenstate components with often non-negligible system-specific fluctuations  at a local scale. Also indicated by previous studies on many body systems, a maximally entangled state displays  ergodic dynamics in the single body product basis. In contrast, a separable state tends to localize in this basis. Between these two extremes,  lies a whole range of states i.e non-maximally entangled states with non-ergodic dynamics. Indeed an ensemble of such states appears in general in any practical quantum information process; this makes it relevant to determine the distribution of the entanglement measures over  the ensemble for a given set of system conditions. The information is desirable e.g. to determine an appropriate purification procedure, or whether a controlled  variation of the system conditions can lead to an ensemble of fully entangled states.

For a clear exposition of our ideas, the present study is confined to the cases with $2nd$ rank state tensor of a pure bipartite state, hereafter referred as the state matrix or just $C$-matrix. We note while the fundamental aspects of the entanglement e.g. many body localization  studies require insights  into multipartite entanglement, its application aspects e.g. security of quantum communication is often based on a shared state between two parties. This  encourages us to confine the present analysis to bipartite entanglement.

The standard entanglement measures for a pure  bipartite state can in principle be obtained from the eigenvalues of the reduced  density matrix. 
Previous studies in this context have been confined to ergodic states; the related reduced density matrix  can then be well-modelled by the basis invariant Wishart random matrix ensembles \cite{page1993average,bianchi2022volume,majumdar2010extreme,kumar2011entanglement,nadal2011statistical,vivo2016random,hall1998random,giraud2007purity,wei2020skewness,li2021moments,huang2021kurtosis,PhysRevB.93.134201}. But, as mentioned above, a typical many-body state,  need not be maximally entangled and can also exhibit different degree of localization \textit{e.g.,} fully or partially localized or extended based on system parameters like disorder strength and dimensionality. The reduced density matrix for such cases belongs to a general class of Wishart random matrix ensembles where system-dependence appears through distribution parameters of the matrix elements \cite{Shukla_2017}. This in turn affects the distribution of the Schmidt eigenvalues and can thereby  cause fluctuations in the entanglement measures  too, an aspect which has been neglected so far to best of our knowledge.  As shown in a previous study \cite{Shekhar_2023}, the probability density of the Schmidt eigenvalues undergoes a diffusive dynamics  with changing system condition, with its variation governed by a measure referred as complexity parameter. In the present work, we apply the formulation to derive the probability densities of the entanglement measures and their growth with changing system conditions. 

The paper is organized as follows.  In  \cite{Shekhar_2023}, we derived the complexity parameter formulation of  the probability density of the Schmidt eigenvalues of a bipartite pure state represented by a multiparametric Wishart ensemble.  As the information is needed to derive the distributions of the entanglement entropies,  it is briefly reviewed  in section \ref{secBasic}. Sections \ref{secS2}  and \ref{secR1} contain the probability density formulation for  the moments of the Schmidt eigenvalues, with primary focus on the purity,  and Von Neumann entropy,  respectively.  For purpose of  clarity, the details of the derivations are presented in the appendices. The numerical analysis presented in section \ref{secNumerics} verifies our theoretical predictions.  We conclude in section \ref{secConc} with a brief summary of our results, their relevance and open questions.

\section{ State matrix ensemble for a pure state} \label{secBasic}

An arbitrary  pure state $|\Psi \rangle$ of a composite system, consisting of two subsystems $A$ and $B$  can be written as $|\Psi \rangle = \sum_{i,j}  C_{ij}  \; |a_i\rangle \; |b_j \rangle$, with
$|a_i\rangle$ and $|b_j \rangle$, $i=1 \ldots N_a, j=1 \ldots N_b$ as the orthogonal basis states in the subspaces of $A$ and $B$ respectively, forming a $N_a \times N_b$ product basis  $ |a_i\rangle \; |b_j \rangle$.

The reduced density matrix $\rho_A$ can then be obtained from the density operator $\rho = |\Psi \rangle \langle \Psi |$ by a partial trace operation on $B$-subspace, $\rho_A   \equiv C\,C^{\dagger}$, subjected to a fixed trace constraint ${\rm Tr}\rho_A =1$. As $C$ consists of the coefficients (components) of the state $\Psi$, we refer it as the state matrix or just $C$-matrix.  A knowledge of the eigenvalues $\lambda_n$ with $n=1 \ldots N_a$ of $\rho_A$, also known as Schmidt eigenvalues,  leads to various entanglement entropies, defined as

\begin{eqnarray}
R_{\alpha} &=& {1\over 1-\alpha} {\rm ln} \; {\rm Tr} (\rho_A ^{\alpha})=  {1\over 1-\alpha} {\rm ln} \; \sum_n \lambda_n^{\alpha}. 
\label{re}
\end{eqnarray}
with the eigenvalues subjected to the trace constraint $\sum_{n=1}^N \lambda_n=1$, $\alpha \to 1$ referring to the  Von Neumann entropy and $\alpha >1$ as the R\'enyi entropies. 

A separable pure  state of a bipartite system can in general be written as 
$|\Psi \rangle = |\phi_a\rangle \; |\phi_b\rangle$ with $|\phi_a \rangle= \sum_{i} a_i \; |a_i\rangle$ and $|\phi_b \rangle= \sum_{j} b_j \; |b_j\rangle$; a typical element of $C$ matrix can then be written as $C_{ij}=a_i b_j$.   This in turn gives  only one of the Schmidt eigenvalues as one and all others zero (due to unit trace constraint) and thereby leads to $R_{alpha}=0$. On the contrary, an ergodic pure state corresponds to almost all $C_{ij}$ of the same order and thereby  corresponding Schmidt eigenvalues of the same order: $\lambda_n \sim {1\over N}$ for $n=1 \ldots N$ again subjected to constraint $\sum_{n=1}^N {\lambda_n}=1$.  This in turn leads to $R_{\alpha} \sim \log N$ \cite{page1993average}. The matrix elements of a $C$ matrix for a generic non-ergodic state in general lie between the two limits.

As mentioned in section \ref{secIntro}, the transmission of a state through a noisy quantum channel can cause randomization of its components. For example, in quantum error correction codes, the state is usually coupled with a few known ancilla states resulting  in a multipartite state. It is then made to pass through a noisy channel, resulting in a randomization of its components. Even if the original state is a non-random one, the state emerging from the noisy channels is best described by an ensemble. This can be further explained by following prototypical example (later referred as example $1$). Consider a pure qubit state  in the computational basis i.e $|\Psi \rangle = \alpha |0\rangle + \gamma |1\rangle$. It passes through a noisy channel described by a unitary operator $U=\begin{pmatrix}a & b  \\ c & d \end{pmatrix}$ with $a, b, c, d$ Gaussian distributed with zero mean and variances $\sigma_a^2, \sigma_b^2, \sigma_c^2, \sigma_d^2$ respectively. The new state $|\Psi' \rangle = U  |\Psi \rangle$ can be written as $|\Psi' \rangle = \alpha' |0\rangle  + \gamma' |1\rangle$. With $\alpha'=a \alpha + b \gamma, \gamma'=c \alpha + d \gamma$ and $a, b, c, d$ as random variables,  this results in $|\Psi' \rangle$ as a random state, with its components $\alpha', \gamma'$ as Gaussian distributed:  
\begin{eqnarray}
\rho_f(\alpha') &=& {1\over 2 \pi \sigma_a \sigma_b } \int  {\rm d}a \, {\rm d}b \, \delta(\alpha'-a \alpha - b \gamma) \; {\rm e}^{-{a^2 \over 2 \sigma_a^2}-{b^2 \over 2 \sigma_b^2}}  \\
&=& { \sigma_a \sigma_b \alpha \over  \sqrt{\pi^3} (\alpha^2 \sigma_a^2 + \gamma^2 \sigma_b^2)} \; {\rm exp} \left[-{\alpha'^2 \over 2 \sigma_a^2 \alpha^2}\left(1-{\alpha^2 \sigma_a^2 \gamma^2 \sigma_b^2 \over \alpha^2 \sigma_a^2 + \gamma^2 \sigma_b^2}\right)\right]
\label{a1}
\end{eqnarray}
 and 
\begin{eqnarray}
\rho_f(\gamma') &=& {1\over 2 \pi \sigma_c \sigma_d } \int {\rm d}c \, {\rm d}d \, \delta(\gamma'-c \alpha - d \gamma) \; {\rm e}^{-{c^2 \over 2 \sigma_c^2}-{d^2 \over 2 \sigma_d^2}} \\
&=& { \sigma_c \sigma_d \alpha \over  \sqrt{\pi^3} (\alpha^2 \sigma_c^2 + \gamma^2 \sigma_d^2)} \; {\rm exp} \left[-{\gamma'^2 \over 2 \sigma_c^2 \alpha^2}\left(1-{\alpha^2 \sigma_c^2 \gamma^2 \sigma_d^2 \over \alpha^2 \sigma_c^2 + \gamma^2 \sigma_d^2}\right)\right]
\label{a2}
\end{eqnarray}
 The state $|\Psi' \rangle$  is now described by the ensemble $\rho_f= \rho_f(\alpha') \rho_f(\gamma')$. We note as 
the components of the state $|\Psi\rangle$ were chosen to be non-random, it is described by a probability density $\rho_i= \delta(\alpha'-\alpha) \delta(\gamma'-\gamma)$. The above gives rise to the query whether it is possible to change the variances $\sigma_a^2, \sigma_b^2, \sigma_c^2, \sigma_d^2$ in time $t$ such that a typical state in the ensemble of states $|\Psi'\rangle$ turns out to be a maximally entangled one? 
Taking $U$ as a $2^L \times 2^L$ random unitary matrix, the above example can also be generalized to show the randomization of a pure bipartite state of   $L$ qubits. Clearly a retrieval of the original information content of the state requires a detailed statistical analysis of the ensemble.

 The randomization of the components of an eigenstate is not confined only to 
environmental effects. A lack of detailed information about the components  (some or all)  of a generic many body state  in a physically motivated basis often leaves them determined only by a distribution. As a consequence, the $C$-matrix for the pure state is best represented by an ensemble, with latter's details subjected to the physical constraints on the state, \textit{e.g.,}  symmetry, conservation laws, dimensionality etc. 
We emphasize, contrary to a mixed state,  the term "ensemble" used in our analysis refers to the one arising due to a randomization of the components of a quantum state $|\Psi \rangle$ due to complexity e.g disorder or many body interactions (e.g. a determination of the components in the product basis requires multiple integrations, leading to approximations). Our analysis is confined to pure states and does not include mixed states where the notion of the ensemble appears due to experimental errors or thermal fluctuations.

We consider the pure bipartite states for which a typical $C$-matrix element is known only by its average and variance.  Following maximum entropy hypothesis,  the $C$-matrix ensemble  can then be described by a multiparametric Gaussian ensemble \cite{Shekhar_2023}
\begin{eqnarray}
{\rho_c}(C; h,b)    &=&  \mathcal{N} \;  {\rm exp}\left[-  \sum_{k,l,s} \;{1\over 2 h_{kl;s}} \left( C_{kl;s} -b_{kl;s} \right)^2 \right] 
\label{rhoc}
\end{eqnarray}
with $\sum_{k,l,s}  \equiv \sum_{k=1}^N \sum_{l=1}^{N_{\nu}} \sum_{s=1}^{\beta}$ and  $\mathcal{N} $ as a normalization constant: $\mathcal{N} = \prod_{k,l, s} (2 \pi h_{kl;s})^{-1/2} $. Here $\beta=1, 2$ for real and complex matrices respectively. Also, $h \equiv \left[h_{kl,s} \right]$ and $b\equiv \left[b_{kl,s} \right]$ refer to the matrices of variances and mean values of $C_{kl;s}$. As the ensemble parameters are governed by correlations among the basis states, different choices of $h$ and $b$-matrices in eq.(\ref{rhoc})  corresponds to the  ensembles representing different pure states.
For example, the ensemble for a separable state with $C_{i1}= a_i b_1$, $C_{ij}=0$ for $j>1$ (representing the state of part $B$ localized to just one local basis state \textit{i.e.,} $|\phi_b \rangle=|b_1\rangle$) can be modelled by eq.(\ref{rhoc}) by choosing $h_{kl;s}, b_{kl;s} \to 0$ for $l \not=1$. The ensemble of pure ergodic states  can similarly be modelled by all $h_{kl;s}$ of the same order and all $b_{kl;s}$ too \textit{e.g.,}   $h_{kl;s} \to \sigma^2$, $b_{kl;s} \to 0$. A choice of varied combinations of $h_{kl;s}, b_{kl;s}$ in general leads to the ensemble of non-ergodic states; some of them are used later in section \ref{secNumerics} for numerical verification of our results.

As discussed in \cite{Shekhar_2023}, a variation of system parameters, \textit{e.g.,}  due to some external perturbation can lead to a dynamics  in the matrix space as well as in the ensemble space, former by a variation of the matrix elements and the latter by a variation of the ensemble parameters. Interestingly, following an exact route,  a specific type of multi-parametric dynamics in the ensemble space (\textit{i.e.,} a particular combination of the first order parametric derivatives of $\rho(H)$) can then be mapped to a Brownian dynamics of $\rho(H)$ in the matrix space. This in turn leads to  complexity parameter   
governed variation of the joint probability distribution function (JPDF)  of the Schmidt eigenvalues  (denoted as $P_{c}(\lambda)$ hereafter) 
\begin{eqnarray}
P_c(\lambda_1, \lambda_2, \ldots,  \lambda_N; Y) =\delta\left(\sum_n \lambda_n -1\right) \; P_{\lambda}(\lambda; Y)
\label{pc}
\end{eqnarray}

where $P(\lambda_1, \lambda_2, \ldots,  \lambda_N; Y)$, referred as $P_{\lambda}(\lambda; Y)$ hereafter,  satisfies following diffusion equation,

\begin{eqnarray}
\frac{\partial P_{\lambda}}{\partial Y}=4 \sum_{n=1}^N \left[\frac{\partial^2 (\lambda_n \; P_{\lambda})}{\partial \lambda_n^2} - \frac{\partial}{\partial \lambda_n}\left( \sum_{m=1}^N \frac{\beta \lambda_n}{\lambda_n- \lambda_m} 
+ {\beta \nu} - { 2 \gamma } \lambda_n  \right)  P_{\lambda}\right] 
\label{pdl1}
\end{eqnarray}
where $\nu=(N_{\nu} - N_a + 1)/2$ and $Y$ is known as the ensemble complexity parameter \cite{shukla2005level,Shukla_2017,PhysRevE.102.032131,shukla2000alternative}
\begin{eqnarray}
Y=-\frac{1}{2 M\gamma} \; {\rm ln} \left[\sideset{}{^{\prime}}{\prod}_{k,l} \prod_{s=1}^{\beta} |(1- 2 \gamma h_{kl;s})  | \;|b_{kl;s}|^2\right]+{\rm const}
\label{y}
\end{eqnarray}
with $\prod_{k,l}' $  implies a product over non-zero $b_{kl;s}$ as well as $h_{kl;s}$, with $M$ as their total number; (for example for the  case with all  $h_{kl;s} \not= {1\over 2 \gamma}$ but $b_{kl;s}=0$, we have $M=\beta N N_{\nu}$ and for case with all $h_{kl;s} \not= {1\over 2 \gamma}$  and $b_{kl;s} \not=0$, we have $M=2\beta N N_{\nu}$). Here  $\gamma$ is an arbitrary parameter, related to final state of the ensemble  (giving the variance of  matrix elements at the end of the variation) and  the constant in eq.(\ref{y}) is determined by the initial state of the ensemble \cite{Shekhar_2024}. As $Y$ in eq.(\ref{pdl1}) appears as a time like variable, hereafter we refer the variation of $P_{\lambda}$ described by the equation as an evolution or growth with $Y$ as a "pseudo-time". 

The above equation describes the diffusion of $P_{\lambda}(\lambda,Y)$, with a finite drift, from an arbitrary initial state $P_{\lambda}(\lambda,Y_0)$ at $Y=Y_0$. For example, if the initial ensemble $\rho_c(C)$ corresponds to that of separable quantum  states, we have
\begin{eqnarray}
P_{\lambda}(\lambda,Y_0) = \sum_{n=1}^N {\rm e}^{-{(\lambda_n-1)^2\over 2 \, \sigma^2}} \,  \prod_{m\not=n} \delta(\lambda_m)
\label{psep}
\end{eqnarray}

In limit $\frac{\partial P_{\lambda}}{\partial Y} \rightarrow 0$ or $Y\rightarrow \infty$, the diffusion approaches a unique steady state:
\begin{eqnarray}
P_{\lambda}(\lambda; \infty) \propto \prod_{m < n=1}^N |\lambda_m -\lambda_n|^{\beta} \; \prod_{k=1}^N |\lambda_k|^{2 \nu \beta -1} \;  {\rm e}^{-{\gamma \over 2 } \sum_{k=1}^N \lambda_k}
\label{perg}
\end{eqnarray}
An important point  worth noting here is as follows:  although the above distribution along with eq.(\ref{pc}) corresponds to the  Schmidt eigenvalues of an ergodic state within unit trace constraint, it does not represent that of a maximally entangled state. In the latter case with Gaussian randomness, we have
\begin{eqnarray}
P_{\lambda}(\lambda,\infty) \propto   {\rm e}^{-{\gamma \over 2 } \sum_{n=1}^N \left(\lambda_n-{\alpha_n\over N} \right)^2}
\label{pmax}
\end{eqnarray}
with $\sum_{n} \alpha_n =N$; an example of $\alpha_n$ satisfying the constraint  in large $N$ limit is $\alpha_n=1$.
Clearly their repulsion  in eq.(\ref{perg}) renders it impossible for the eigenvalues to reach to the same value $1/N$. Further we note that while all but one Schmidt eigenvalues of a separable state cluster around zero, those for maximally entangled state cluster around $1/N$. Thus while the quantum state itself may change from localized to ergodic limit, the Schmidt states \textit{i.e.,} eigenstates of $\rho_A$ seem to behave differently. But as the entanglement entropy for an ergodic state is almost maximum, this suggests different routes for maximum entanglement for random and non-random states.

As discussed in detail in \cite{Shekhar_2023, Shekhar_2024}, the parametric variation described in eq.(\ref{pdl1}) is subject to a set of constants of dynamics. The latter arise from a mapping of the set $\{h,b\}$ of ensemble parameters (say $M$ of them varying) to another set $\{Y, Y_2, \ldots, Y_M \}$ so that the first order partial derivatives with respect to  $M$ ensemble parameters reduces to a single partial derivative with respect to $Y$. The remaining parameters $Y_2, \ldots, Y_M$ play the role of constants of dynamics. As the relevance of $Y_2, \ldots Y_M$ has been discussed in detail \cite{Shekhar_2023, Shekhar_2024} and explained through examples in \cite{Bera2016}, we discuss only the role of $Y$  in the present analysis.

\section{ Distribution of the moments of Schmidt eigenvalues} \label{secS2}

As eq.(\ref{re}) indicates, the R\'enyi entropy $R_k$ is  related to the moments $S_k (\equiv \sum_n \lambda_n^k)$ of the Schmidt eigenvalues, and it is  appropriate to first  consider the probability density of $S_k$, $k >1$, defined as 
\begin{eqnarray}
f_k(S_k; Y) = \int \delta\left(S_k-\sum_{n=1}^N \lambda_n^k\right) \;\; \delta\left(\sum_{n=1}^N \lambda_n -1\right) \; P_{\lambda}(\lambda; Y) \; {\rm D}\lambda 
\label{fk}
\end{eqnarray}
Using the product of $\delta$-functions in the above integral, 
$f_k$ can alternatively be defined in terms of the joint probability density  of the moments $P_q(S_1, S_2,\ldots, S_q) $ 
\begin{eqnarray}
f_k(S_k; Y) = \int \delta(S_1-1) \; P_q(S_1,\ldots, S_q; Y) \; \prod_{m=1 \atop m \not=k}^q {\rm d}S_k.
\label{fk1}
\end{eqnarray}
where 
 \begin{eqnarray}
P_q(S_1, S_2,\ldots, S_q; Y) 
=\int \prod_{k=1}^q\delta(S_k-\sum_{n=1}^N \lambda_n^k) \; P_{\lambda}(\lambda; Y) \; {\rm D}\lambda 
\label{psq}
\end{eqnarray}

For clarity purposes, here the notation $f_k(S_k)$ is reserved for the distribution of a single moment $S_n$ and  $P_q$ to the JPDF of $q$ such moments.
Also, to avoid cluttering of presentation, henceforth we use following notations interchangeably: $\sum_n \equiv \sum_{n=1}^N$, $\delta\left(\sum_{n=1}^N \lambda_n -1\right) \equiv \delta_1$ and $\delta_{S_k} \equiv \delta\left(S_k-\sum_{n=1}^N \lambda_n^k\right) $. Also, for simplicity, hereafter we consider the case of $k=2$. 

\subsection{Distribution of purity} 

From eq.(\ref{fk1}), the probability density of purity $S_2$   can be written as
\begin{eqnarray}
f_2(S_2; Y) = \int_0^{\infty} \delta(S_1-1) \; P_2(S_1, S_2; Y) \; {\rm d}S_1,
\label{f2}
\end{eqnarray}
where, $S_1 \equiv \sum_n \lambda_n$. We note that the $\delta$-unction in the above acts like a filter, picking  the value of $ P_2(S_1, S_2; Y)$ at $S_1=1$. It can therefore be replaced by a narrow width function $g(S_1)$ centred at $S_1=1$ such that (i) the product $g(S_1) P_2(S_1, S_2; Y)$ is non-zero only in the vicinity of $S_1=1$, and, (ii) integrating the product over $S_1$  gives $g(S_1) P_2(1, S_2; Y)$. Using the definition $f_2(S_2; Y) =f_{2, \omega}(S_2; Y)$ in large $\omega$ limit, eq.(\ref{f2}) can now be rewritten as 

\begin{eqnarray}
f_{2, \omega}(S_2; Y) = \int_0^{\infty}  g(S_1) \; P_2(S_1, S_2; Y) \; {\rm d}S_1 
\label{f2w}
\end{eqnarray}
Differentiation of the above equation with respect to $Y$ gives the $Y$-dependent growth of $f_{2, \omega}$:
\begin{eqnarray}
\frac{\partial f_{2, \omega}}{\partial Y} = \int_0^{\infty} \, g(S_1) \; \frac{\partial P_2}{\partial Y} \; {\rm d}S_1.
\label{f2y}
\end{eqnarray}
To proceed further, a knowledge of the $Y$- governed variation of $P_2(S_1, S_2; Y)$ is required; this can be obtained by differentiating eq.(\ref{psq}) for $q=2$ with respect to $Y$ and subsequently using eq.(\ref{pdl1}) (details discussed in {\it appendix} \ref{ps2pvn}):

\begin{eqnarray}
\frac{\partial P_2}{\partial Y}= 4 \frac{\partial^2 (S_2 \; P_2)}{\partial S_2 \partial S_1} + 4 \frac{\partial^2 I_3}{\partial S_2^2} + \frac{\partial^2 (S_1 \; P_2)}{\partial S_1^2} + \frac{\partial (Q_2 \; P_2)}{\partial S_2} +\frac{\partial (Q_1 \; P_2)}{\partial S_1}
\label{ps2}
\end{eqnarray}

with $Q_2 \equiv (2 \gamma S_2 - \beta(N + \nu - 1)S_1- S_1)$, $Q_1= 2 \gamma  S_1-{\beta\over 2} N (N+2 \nu-1)$ and 

\begin{eqnarray}
I_3 = \int \prod_{k=1}^2 \; \delta_{S_k} \; \left(\sum_n \lambda_n^3 \right) \;  P_{\lambda} \; {\rm D}\lambda = \int \; S_3 \; P_3(S_1, S_2,S_3) \; {\rm d}S_3.
\label{i3}
\end{eqnarray} 
We note that all other terms except one in  eq.(\ref{ps2}) contain the derivatives of $P_2$.  
To reduce it as a closed form equation for $P_2$, we rewrite  $P_3(S_1, S_2,S_3)= P(S_3|S_2, S_1)\; P_2(S_2, S_1)$. This gives $I_3 = \langle S_3 \rangle_{S_2, S_1} \; P_2$ where $\langle S_3 \rangle_{S_2, S_1}= \int \; S_3 \; P(S_3|S_2, S_1) \; {\rm d}S_3$ is the ``local average'' \textit{i.e.,} ensemble average of $S_3$ for a given $S_2, S_1$. Substitution of the so obtained  $I_3$ form in eq.(\ref{ps2})  gives now the diffusion equation for $P_2$ for arbitrary $S_1$ in a closed form.

Using eq.(\ref{ps2}) in eq.(\ref{f2y}), integrating by parts, and subsequently using the relation $f_{2, \omega}(S_2) \to 0$ at the integration limits $S_1=0, \infty$,  now leads to

\begin{eqnarray}
\frac{\partial f_{2, \omega}}{\partial Y}=  \frac{\partial^2 f_{2, \omega}}{\partial S_2^2} +(\eta \; S_2 + b) \frac{\partial f_{2, \omega}}{\partial S_2}  + d_0 \; f_{2, \omega},
\label{pf}
\end{eqnarray}
with,
\begin{eqnarray}
a &=& 4  \langle S_3 \rangle_{S_2, S_1=1}, \\
b &=& -\beta(N+ \nu-1) + 2 \frac{\partial a}{\partial S_2},\\
\eta &=&  2 \gamma - 4 \; g_1, \\
d_0 &=&  4 \gamma - (6+ 2 \gamma -{\beta \over 2} N N_{\nu}) g_1 + g_2 + \frac{\partial^2 a}{\partial S_2^2},
\label{abcd}
\end{eqnarray}
where, $ g_n(S_1) \equiv {1\over g} {\partial^n g(S_1)\over \partial S_1^n}$ and $N_{\nu}=N+2 \nu-1$.   Here again the above closed form equation is obtained by approximating the product term $g(S_1) \; \langle S_3 \rangle_{S_2, S_1}$ in the following integrand by its value at $S_1=1$ (the peak of $g(S_1)$),
\begin{eqnarray}
\int_0^{\infty}  g(S_1) \; \langle S_3 \rangle_{S_2, S_1} \; P_2(S_1, S_2; Y) \; {\rm d}S_1 = \langle S_3 \rangle_{S_2, S_1=1}
\label{inta}
\end{eqnarray}

We note $\langle S_3 \rangle$  is in general a function of $Y$, leaving  the right side of eq.(\ref{pf}) $Y$-dependent. This  makes it technically complicated to solve; further progress can however be made by noting that   $ S_3$  and $S_2$ change at different rates with $Y$, with $S_3$ changing much slower than $S_2$ and   can be assumed almost constant as $S_2$ evolves with $Y$.  This permits us to replace $\langle S_3 \rangle_{S_2, S_1=1}$  by the latter averaged over all $S_2$, referred as $\langle S_3 \rangle$, ignore terms $\frac{\partial a}{\partial S_2}$ and $\frac{\partial^2 a}{\partial S_2^2}$  in eq.(\ref{abcd}) and consider a separation of variables approach to  solve eq.(\ref{pf}).

Although in principle it is possible to proceed 
without considering any specific form of the limiting function  $g(S_1)$,   for technical clarity here we  consider

\begin{eqnarray}
g(S_1)=\omega \; {\rm e}^{-\omega \, (S_1-1)}.
\label{g1}
\end{eqnarray}
While the above function  approaches a $\delta$-function (for $S_1 \ge 1$) in limit $\omega \to \infty$), but to act as a desired filter, it is sufficient to have its width  much smaller than that of $P_2(S_1, S_2; Y)$. This in turn gives $ g_n(S_1) = (-1)^n \, \omega^n$ and thereby $\eta \approx 4 \omega$ (in large $\omega$-limit and as $\gamma$ a fixed constant).
Further, with our primary interest in $S_1=1$, it is sufficient to consider the above form instead of the form $\omega \; {\rm e}^{-\omega \, |S_1-1|}$ valid for entire real line.  
  
A general solution  of the above equation then gives the growth of the probability density of purity from arbitrary initial condition (at $Y=Y_0$) as  $Y$ varies. But, with $\langle S_3 \rangle$ in general  $Y$-dependent, the above equation is non-linear in $Y$ and is technically difficult to solve. Further steps can however be simplified by
considering a rescaled variable $x$ defined as 

\begin{eqnarray}
x = \left({\omega\over 2 \, \langle S_3 \rangle} \right)^{1\over 2} \, S_2.
\label{xap}
\end{eqnarray}
Further defining $\Psi(x, Y)$ as the transformed probability,  given by the relation $f_{2, \omega}(S_2, Y) = \Psi(x, Y)  {{\rm d} x \over {\rm d}S_2}$,  eq.(\ref{pf}) can now be rewritten as 

\begin{eqnarray}
\frac{\partial \Psi}{\partial Y}= 2 \omega \; \frac{\partial^2 \Psi}{\partial x^2} +\eta \, (x + b_1 ) \frac{\partial \Psi}{\partial x}  + d_0 \; \Psi
\label{ppsi}
\end{eqnarray}
where $b_1= {b\over \eta} \sqrt{ \omega \over 2 \,  \langle S_3 \rangle}$.  Further, since  $\eta \approx  4 \omega$, we can approximate  $x+b_1  \approx x$, thereby leaving the right side of the above equation $Y$-independent. 
Eq.(\ref{ppsi}) can now be solved by the separation of variables approach;  as discussed in  {\it appendix} \ref{solns2} in detail, the general solution  for arbitrary $Y-Y_0$ can be given as 

\begin{eqnarray}
\Psi(x; Y-Y_0) &\approx &   {\rm e}^{-{x^2}}  \; \sum_{m=0}^{\infty} \, C_{m}  \;  _1F_1\left(-{\mu_m}, {1\over 2},  x^2 \right) \; {\rm e}^{- d_0 m \; (Y-Y_0)}\; 
\label{ps8} 
\end{eqnarray}
where, $_1F_1(a,b,z)$ is the \textit{Kummer's} Hypergeometric function \cite{dlmf} with
\begin{eqnarray}
\mu_m = \mu_0 \, (m + 1),  \qquad \mu_0 = {d_0\over 8 \, \omega} \approx  {1\over 16} (2 \omega - \beta N N_{\nu} ).
\label{mum}
\end{eqnarray}

Substitution of eq.(\ref{ps8}) in the relation 
\begin{eqnarray}
f_{2, \omega}(S_2, Y) = \Psi(x,  Y) \sqrt{\omega \over 2 \langle S_3 \rangle}
\label{fpsi}
\end{eqnarray}
leads to $f_{2, \omega}(S_2, Y)$. We emphasize  that the solution in eq.(\ref{ps8}) is applicable for arbitrary $N$. The only approximation used is eq.(\ref{inta}) which was used to derive a closed form of  eq.(\ref{pf}).

{\bf Limit $(Y-Y_0) \to \infty$:} As a check, we  consider  the solution in $Y-Y_0 \to \infty$ limit. In this case, the non-zero contribution comes only from the term $m=0$ in eq.(\ref{ps8}), giving 

\begin{eqnarray}
\Psi(x; \infty) &\approx & C_0 \;   {\rm e}^{-{x^2}} \, _1F_1\left(-\mu_0, {1\over 2},  x^2 \right). 
\label{ps10}
\end{eqnarray}

Substitution of  eq.(\ref{ps10}) in eq.(\ref{fpsi}) gives

\begin{eqnarray}
f_2(S_2; \infty) &\approx & C_0 \, \sqrt{\omega \over 2 \langle S_3 \rangle} \;  {\rm e}^{-{\omega S_2^2 \over 2 \langle S_3 \rangle}} \, _1F_1\left(-{\mu_0\over 2}, {1\over 2},  {\omega \, S_2^2 \over 2 \langle S_3 \rangle} \right).
\label{finf}
\end{eqnarray}
Further, with $\omega > N^2$, the limit $N \to \infty$ implies $\omega \to \infty$. Using the limiting Gaussian form of a $\delta$-function, we now have 
\begin{eqnarray}
f_2(S_2; \infty) 
\to \delta(S_2)
\label{finf1}
\end{eqnarray}
which is consistent with large $N$-limit of $S_2 \sim 1/N$ in the entangled limit.

{\bf Finite $(Y-Y_0)$:} 
For a finite $N$, $\omega$  can be chosen large but finite.
A smooth transition for finite $N$ can then be seen in terms of the rescaled variable $x$ (eq.(\ref{xap})) and a rescaled evolution parameter $\Lambda= 8 \omega \mu_0 (Y-Y_0)$,

\begin{eqnarray}
\Psi(x; Y-Y_0) &\approx &   {\rm e}^{-{x^2}}  \; \sum_{m=0}^{\infty} \, C_{m}  \;  _1F_1\left(-{\mu_m}, {1\over 2},  x^2 \right) \; {\rm e}^{- m  \, \Lambda}
\label{ps11} 
\end{eqnarray}
Here, from eq.(\ref{mum}), $\mu_0$ is  positive and finite with $\omega > (\beta/2) N N_{\nu}$ for any arbitrary $N$.

The unknown constants $C_m$ in the above equation can be determined if the initial distribution  $\Psi(x; Y_0)$ is known. Further insight can however be 
gained directly by using  the expansion

\begin{eqnarray}
_1F_1\left(-{\mu_m}, {1\over 2},  x^2 \right) =  _1F_1\left(-{\mu_0}, {1\over 2},  x^2 \right) \; \left( 1 - 2 m \mu_0 x^2 +  q_{1m} \, x^4 + O(x^6) \right)
\label{hyp2}
\end{eqnarray}
 with $q_{1m} \equiv {1\over 3} m \mu_0 (\mu_0(m-4)-1)$. Near $x=0$, eq.(\ref{ps11}) can now be rewritten as
 
\begin{eqnarray}
\Psi(x; Y-Y_0) &\approx &   \Psi(x; \infty)  \; \sum_{m=0}^{\infty} {C_m \, {\rm e}^{-m \Lambda}\over C_0}  \; \left( 1 - 2 m \mu_0 x^2 + q_{1m} \, x^4 + O(x^6) \right)  
\label{ps12} 
\end{eqnarray}
For large $\Lambda$, only lower $m$ orders contribute to the series for arbitrary $x$, giving

\begin{eqnarray}
\Psi(x; Y-Y_0) &\approx &   \Psi(x; \infty) \, \left[1 + {C_1 \, {\rm e}^{-\Lambda}\over C_0} \left( 1 - 2 \mu_0 x^2 + q_{11} \, x^4 + O(x^6) \right)  \right],  \hspace{0.3in}  x \sim 0 \label{ps13} \\
 &\approx &  \Psi(x; \infty) +  C_{1}  \;  {\rm e}^{-{x^2}}  \;   _1F_1\left(-2 \mu_0, {1\over 2},  x^2 \right) \; {\rm e}^{-\Lambda}  \hspace{0.3in}  x \; {\rm arbitrary} 
\label{ps14} 
\end{eqnarray}
From eq.(\ref{ps13}), the large $\Lambda$ distribution near $x=0$ is thus predicted to be analogous to  $\Psi(x; \infty)$. For arbitrary $x$, however,  the additional term  in eq.(\ref{ps14})  perturbs $\Psi(x; \infty)$,  changing the distribution parameters slightly  while almost retaining its shape;  This is also indicated by our numerical analysis discussed in next section.

For any small $\Lambda >0$, while the factor $ e^{-m \Lambda}$ in eq.(\ref{ps11})  becomes increasingly small with increasing $m$,  this can be compensated by the large $C_m$ values  and can thereby  lead to an intermediate state for $\Psi(x; Y)$, although different from the one at $Y=Y_0$ but close to it. Writing ${\rm e}^{- m  \, \Lambda} \approx 1 - m \Lambda + O(m^2 \Lambda^2)$, eq.(\ref{ps11}) gives for arbitrary $x$,

\begin{eqnarray}
\Psi(x; Y-Y_0) &\approx &  \Psi(x, 0) - \Lambda \, {\rm e}^{-{x^2}}  \; \sum_{m=0}^{\infty} \, m \, C_{m}  \;  _1F_1\left(-{\mu_m}, {1\over 2},  x^2 \right) 
\label{ps13} 
\end{eqnarray}
For $x \sim 0$, and, with help of eq.(\ref{hyp2}), the above can be rewritten as

\begin{eqnarray}
\Psi(x; Y-Y_0) &\approx &  \Psi(x, 0) - \Lambda \, \Psi(x, \infty) \;   
\sum_{m=0}^{\infty} {m C_m \over C_0}  \; \left( 1 - 2 m \mu_0 x^2 + q_{1m} \, x^4 + O(x^6) \right)  
\label{ps14} 
\end{eqnarray}
The above indicates the  distribution near $x \sim 0$ for finite $Y$ as a
superposition of the two distributions: the initial one at $Y=Y_0$ and the other as  a perturbed distribution at $Y\to \infty$.

The limit $N\to\infty$ makes it necessary to consider the limit $\omega\to\infty$; a better insight in this case can be derived by writing the distribution in terms of $S_2$ instead of $x$. Using eq.(\ref{ps11}) in eq.(\ref{fpsi}) and large order approximation of the function $_1F_1\left(-{\mu_m}, {1\over 2},  x^2 \right)$ (given in  {\it appendix} \ref{solns2})), we have

\begin{eqnarray}
f_{2, \omega}(S_2, Y-Y_0) &\approx & {2 \;  {\rm e}^{-{\omega \, S_2^2 \over 4 \, \langle S_3 \rangle}}  \over \sqrt{\pi \,\langle S_3 \rangle}}  \,  \sum_{m=0}^{\infty} \; {C_{m}\over (m+1)}   \; \cos\left(\sqrt{(m+1)  \omega^2 \, S_2^2 \over 8 \, \langle S_3 \rangle}\right) \;  {\rm e}^{- 8 \, m \, \omega \, \mu_0 \; (Y-Y_0)}.
\label{fs9}
\end{eqnarray}
As clear from the above, for finite $(Y-Y_0) > 0$,  the terms with $m > 0$ can then contribute significantly only if $\omega \mu_0 (Y-Y_0) \ll  1$ or the coefficients $C_m$ in eq.(\ref{ps8}) are exponentially large overcoming the decay due to ${\rm e}^{- 8 m \; \omega \mu_0 \, (Y-Y_0)}$. As a generic initial state can not lead to such coefficients, we have
\begin{eqnarray}
f_{2, \omega}(S_2, Y-Y_0) &\approx & {2 \; C_0   \over \sqrt{\pi \,\langle S_3 \rangle}}  \,  {\rm e}^{-{\omega \, S_2^2 \over 4 \, \langle S_3 \rangle}}   \; \cos\left(\sqrt{\omega^2 \, S_2^2 \over 8 \, \langle S_3 \rangle}\right).
\label{fs10}
\end{eqnarray}
The above suggests an abrupt transition, in limit $N \to \infty$,  from an initial distribution at $Y-Y_0=0$ to the maximum purity for $Y-Y_0 >0$.

As eq.(\ref{ps8}) indicates,  a knowledge of the average purity of a typical bipartite eigenstate, in a finite Hilbert space, with corresponding state ensemble given by eq.(\ref{rhoc})  is  not enough; its fluctuations are also important. This can also be seen by a direct calculation of the $Y$-dependent growth of the variance of purity  defined as $\sigma^2(S_2) = \langle \Delta S_2^2 \rangle =\langle (S_2)^2 \rangle - \langle S_2\rangle^2$ with $\langle (S_2)^n \rangle = \int  S_2^n \; f_2(S_2;Y) \;  {\rm d}S_2$); eq.(\ref{pf}) gives 

\begin{eqnarray}
\sigma^2(S_2)  \approx  \frac{\langle S_3 \rangle}{\omega} - A \, e^{-8 \Lambda}
\label{sigs2}
\end{eqnarray}
with $A$ dependent on the initial state $Y=Y_0$ (details discussed in {\it appendix} \ref{vars2}). Clearly the variance tends to zero for any  finite $Y-Y_0$ in  limit $\omega, N \to \infty$. The distribution  however retains a finite width for  finite $\Lambda$ equivalently finite $N$ as well as $Y-Y_0$.

The relation $R_2= -\log S_2$ along with eq.(\ref{ps8})   can further be used to derive the probability density of second R\'enyi entropy $R_2$.

\vspace{0.1in}

\subsection{ Joint Probability Distribution of all moments}
For higher order entropies,  it is technically easier to first derive  the  diffusion equation for 
\begin{eqnarray}
P_{\infty}(S_1, S_2, \ldots, S_{\infty}) = \int \prod_{k=1}^{\infty} \delta(S_k-\sum_n \lambda_n^k) \; P_{\lambda} \; {\rm D}\lambda
\label{pinf}
\end{eqnarray}
Following similar steps as in case of $S_2$ but no longer using any approximation, it can be shown that 

\begin{eqnarray}
\frac{\partial P_{\infty}}{\partial Y}=\sum_{q=1}^{\infty} q \;\frac{\partial}{\partial S_q} \left[ \sum_{t=1 \atop t\not=q}^{\infty} t \; \frac{\partial  }{\partial S_t} S_{q+t-1} +  \frac{\partial  }{\partial S_q} S_{2q-1}+ F(q) \right] P_{\infty}
\label{pdsi}
\end{eqnarray}
where $F(q)= 2  S_q +(\eta \beta - (q-1)) S_{q-1} +{\beta\over 2} \sum_{r=0}^{q-1} (S_{q-r-1} S_r -S_{q-1})$.

An integration over undesired variables now leads to the desired distribution. 
For example, $P_q$ (defined in eq.(\ref{psq}) can also be expressed as 
\begin{eqnarray}
P_{q}= \int P_{\infty}(S_1, S_2, \ldots, S_{\infty}) \; \prod_{t=q+1}^{\infty} {\rm d}S_t.
\label{pdsi1}
\end{eqnarray}
The latter expression along with eq.(\ref{pdsi}) then leads to a diffusion equation for arbitrary $q$.  Eq.(\ref{pdsi}) also indicates that the lower order moments of the Schmidt eigenvalues and thereby R\'enyi entropies are dependent on higher order ones. This again indicates that the fluctuations of the entropies can not be ignored.

\vspace{0.2in}

\section{ Distribution of Von Neumann entropy}  \label{secR1}

The diffusion equation for the probability density of the  Von Neumann entropy,   defined in eq.(\ref{re}) for $\alpha=1$,   can similarly be derived.   Using the alternative  definition  $R_1 = -\sum_{n=1}^N \lambda_n \, \log \lambda_n$,   the probability density of $R_1$   can be written as

\begin{eqnarray}
f_v(R_1; Y) = \int_0^{\infty} \delta(S_1-1) \; P_v(R_1, S_1; Y) \; {\rm d}S_1,
\label{fv}
\end{eqnarray}
 where  $P_v(R_1, S_1; Y)$  is the joint probability density of $R_1$ and $S_1$ without the constraint $S_1=1$,
\begin{eqnarray}
P_{v}(R_1, S_1) = \int \delta(R_1+\sum_n \lambda_n \; {\rm log} \lambda_n) \; \delta(S_1-\sum_n \lambda_n) \;  P_{\lambda} \; {\rm D}\lambda.
\label{pv}
\end{eqnarray}

As in the case of purity, $\delta(S_1-1)$ in eq.(\ref{fv}) can again be replaced  by its limiting form $g(S_1)$ (eq.(\ref{g1})),  leading to following distribution

\begin{eqnarray}
f_{v, \omega}(R_1; Y) = \int_0^{\infty} g(S_1) \; P_v(R_1, S_1; Y) \; {\rm d}S_1. 
\label{fvg}
\end{eqnarray}
To proceed further, here again we require a prior knowledge of $\frac{\partial P_v}{\partial Y}$. Differentiating the  integral eq. (\ref{pv}) with respect to $Y,$  and using eq.(\ref{pdl1}) followed by subsequent reduction of the terms by repeated partial integration leads to (see {\it appendix} \ref{ps2pvn}),

\begin{eqnarray}
    \frac{\partial P_v}{\partial Y} &=&  2\frac{\partial^2 }{\partial R_1 \partial S_1} \left[(R_1-S_1) \; P_v\right]+  \frac{\partial^2 }{\partial R_1^2} \left[(S_1-2 R_1)P_v + \langle T_1 \rangle P_v 
    \right]+ \frac{\partial^2 }{\partial S_1^2} (S_1 \; P_v) \nonumber \\
    &+& \frac{\partial}{\partial R_1}\left[\left(\beta N (N + \nu - 1) + 2 \gamma (R_1 - S_1) - \beta \frac{N_{\nu}}{2} \langle R_0 \rangle + N\right) P_v\right] \nonumber \\
    &+& \frac{\partial }{\partial S_1} \left[\left(2  \; \gamma \; S_1-{1\over 2} \beta NN_{\nu}\right)\; P_v \right]
    \label{pvn}
\end{eqnarray}
with $T_k \equiv \sum_n (\lambda_n)^{k} (\log \lambda_n)^{k+1}$. Here as discussed in {\it appendix} \ref{ps2pvn}, the above closed form  equation for $P_v$ is obtained  by using the conditional probability relation $P(R_1,  S_1, T_1) = P(T_1|R_1, S_1)\; P_v(R_1, S_1)$ and writing 

\begin{eqnarray}
\int  T_1 \; P(R_1,  S_1, T_1) \; {\rm d}T_1 = \langle T_1 \rangle_{R_1, S_1} \; P_v(R_1, S_1).
\label{jj1}
\end{eqnarray}
where $\langle T_1 \rangle_{R_1, S_1}$  is the ``local average'' \textit{i.e.,} ensemble average of $T_1$ for a given $R_1, S_1$: $\langle T_1 \rangle_{R_1, S_1}= \int \; T_1 \; P(T_1 |R_1, S_1) \; {\rm d}T_1$.

Differentiating eq. (\ref{fvg}) with respect to $Y$ and subsequently using eq.(\ref{pvn}); a repeated integration by parts along with the relation $f_{v, \omega}(R_1) \to 0$ at the limits $S_1=0, \infty$,   now leads to following diffusion equation 

\begin{eqnarray} 
\frac{\partial f_{v, \omega}}{\partial Y}= \left(t-2 R_1 \right) \frac{\partial^2 f_{v, \omega} }{\partial R_1^2} +\left(a_1 R_1 + b_1 \right) \frac{\partial  f_{v, \omega}}{\partial R_1} + d_0 \;  f_{v, \omega},
\label{pfv}
\end{eqnarray}
with,
\begin{eqnarray}
a_1 &=& 2 \gamma + 2 \, \omega \approx 2 \omega,  \\ 
b_1 &=& \beta  \, (N_{\nu} - \nu) \, N - {\beta \over 2}   N_{\nu} \, \langle R_0 \rangle + N + 2 \, \omega - 2\gamma -4  + 2 \frac{\partial t}{\partial R_1},  \\
d_0 &=& {\beta\over 2} \, \omega \, N \, N_{\nu}  + \omega^2 +
 (2-2\gamma) \; \omega - 2\gamma +\frac{\partial^2 t}{\partial R_1^2}, \\
t &=& 1+\langle T_1 \rangle_{R_1, S_1=1}, \\
\langle R_0 \rangle &=& -\sum_n \langle \log \lambda_n \rangle \approx - N \langle \log \lambda_n \rangle_{e,s}
\label{abcd1}
\end{eqnarray}
where $N_{\nu} = N+2\nu-1$,  and $\langle R_0 \rangle$ implies an ensemble and spectral averaged logarithm of a typical Schmidt eigenvalue (indicated by notation $\langle \rangle_{e,s}$).  Here again the above closed form equation is obtained by following approximation
\begin{eqnarray}
\int_0^{\infty}  g(S_1) \; \langle T_1 \rangle_{R_1, S_1} P_2(R_1, S_1; Y) \; {\rm d}S_1 \approx \langle T_1 \rangle_{R_1, S_1=1}.
\label{int1}
\end{eqnarray}
The above follows by invoking the filtering role of $g(S_1)$; the latter permits  approximation of  the product term $g(S_1) \; \langle T_1 \rangle_{R_1, S_1}$  by its value at $S_1=1$ (the peak of $g(S_1)$).  
Although here, again, both $\langle T_1\rangle$ and $\langle R_0 \rangle$ are a function of $Y$, and consequently the right side of eq.(\ref{pfv}) in this case too depends on $Y$ making it technically complicated to solve; nevertheless, both  $\langle T_1\rangle$ and $\langle R_0 \rangle$, changing much slower than $R_1$,   can be assumed almost constant as $R_1$ evolves with $Y$.  This permits us to replace $\langle T_1 \rangle_{R_1, S_1=1}$ by its $R_1$ averaged value $\langle T_1 \rangle$, ignore terms $\frac{\partial t}{\partial R_1}$ and $\frac{\partial^2 t}{\partial R_1^2}$  in eq.(\ref{abcd1}) and consider a separation of variables approach to  solve eq.(\ref{pfv}).

Again defining $\Psi_v(x, Y)$ as the transformed probability,  given by the relation $f_{v, \omega}(R_1, Y) = \Psi_v(x, Y)  {{\rm d} x \over {\rm d}R_1} $,  the  general solution for an arbitrary $Y-Y_0$ can now  be given as 
({\it appendix} \ref{solnr1}),

\begin{eqnarray}
\Psi_v(x; Y-Y_0) &=&  {1\over 2 \omega}  \left({ 4x\over \omega}\right)^{\alpha}  \; \sum_{m=0}^{\infty} C_{1m} \; {\rm e}^{- \, m |d_0| \, (Y-Y_0)} \;  \; _1F_1\left(\alpha +{ d_m\over 2\omega},    \alpha+1, x \right)  \nonumber \\
\label{pv1}
\end{eqnarray}
with $_1F_1(a,b,x)$ as the Kummer's Hypergeometric function \cite{dlmf}, 
 $\alpha={1\over 4}(a_1 t+2 b_1 +4) = {1\over 2}(\omega t +b_1)$,  $d_m = d_0(m+1)$ and 
\begin{eqnarray}
x \equiv -\omega  (t-2 R_1)
\label{xar}
\end{eqnarray}
Here, following from the definition in eq.(\ref{abcd1}), $t \sim 1+ \langle R_1\rangle^2$.

As in the purity case, here again with $\omega > N^2$   for arbitrary  $N$ (with $\omega$ large but finite), we can approximate $d_0 \approx \omega^2$ and $b_1  \approx 2 \omega$ in eq.(\ref{abcd1}) (with $\langle R_0 \rangle \sim N \log N + N$ \cite{Shekhar_2023}). This in turn gives $\alpha \approx {\omega t/2}$. Further using the identity $_1F_1(a, b;x) = {\rm e}^x \, _1F_1(b-a,  b; -x) $ \cite{dlmf}, we can rewrite eq.(\ref{pv1}) as 

\begin{eqnarray}
\Psi_v(x; Y-Y_0) &=&   {1\over 2 \omega} \, \left({ 4x\over \omega}\right)^{\alpha}  \;  {\rm e}^{x} \; \sum_{m=0}^{\infty} C_{1m} \; {\rm e}^{- \, m \omega^2 \, (Y-Y_0)} \;  {\mathcal F_M} 
\label{pvg}
\end{eqnarray}
where
\begin{eqnarray}
{\mathcal F}_m  &\equiv&  _1F_1\left(1-{\omega (m+1)\over 2}, \, {\omega t\over 2}, \, -x \right) \approx   {\rm e}^{(m+1) x\over t}
\label{pvg1}
\end{eqnarray}
and the unknown constant $C_{1m}$  determined by a prior knowledge of $f_v(R_1, Y_0)$. We note that the solution in eq.(\ref{pv1}) as well as eq.(\ref{pvg}) is applicable for arbitrary $N$; the only approximation considered here is eq.(\ref{jj1}),  used to derive a closed form of  eq.(\ref{pfv}).

{\bf Limit $(Y-Y_0) \to \infty$:} In the limit $Y-Y_0 \to \infty$, the non-zero contribution in eq.(\ref{pvg}) comes only from the term $m=0$, giving 

\begin{eqnarray}
\Psi_{v}(x; \infty) &=&   {C_{10}\over 2 \omega} \, \left({4x\over \omega}\right)^{\omega t/2} \;  {\rm e}^{x}  \;  {\mathcal F}_0 
\label{pv2}
\end{eqnarray} 
 Substitution of eq.(\ref{xar}) in the above equation and using $f_{v, \omega}(x; Y) = 2 \omega \Psi_{v}(x; Y)$, we have 
 
\begin{eqnarray}
f_{v, \omega}(x; \infty) 
&=&  C_{10}  \;  (2R_1-t)^{\omega t/2}  \;  {\rm e}^{- \omega (t-2 R_1)}.
\label{pv3}
\end{eqnarray} 

Using the limit  $f_{v}(R_1; \infty) = \lim_{\omega \to \infty} f_{v, \omega}(R_1; \infty)$,   the above then gives 

\begin{eqnarray}
f_{v}(R_1; \infty) \to \delta(t-2 R_1).
\label{pv4}
\end{eqnarray}

{\bf Finite $(Y-Y_0)$:}  For finite $N$, the condition $\omega > N^2$ can be satisfied by a large but finite $\omega$. A smooth crossover from initial distribution at $Y=Y_0$ to that at $Y \to \infty$ can then be seen in terms of the variable $x$ and rescaled evolution parameter $\Lambda=\omega^2 (Y-Y_0)$,

\begin{eqnarray}
\Psi_{v}(x; Y-Y_0) &=&  {1\over 2 \omega} \, \left({ 4 x\over \omega}\right)^{\omega t/2}  \;  {\rm e}^{{x(t+1)\over t}} \; \sum_{m=0}^{\infty} C_{1m} \;  {\rm e}^{ {m (x - t \, \Lambda)\over t} }
\label{pvg2}
\end{eqnarray}

Using $Y=Y_0$ in eq.(\ref{pvg}), we have 

\begin{eqnarray}
\Psi_{v}(x; 0)  &=&   {1\over 2 \omega} \,  \left({ 4 x \over \omega}\right)^{\omega t/2}   \; {\rm e}^{ {x \, (t+1)\over t}}  \; \sum_{m=0}^{\infty} C_{1m} \;  {\rm e}^{{m x\over t}}.
\label{pvgg}
\end{eqnarray}
We note that right side of the above equation is in the form of a discrete Laplace transform;  with $t$ large,  the sum can be converted into a continuous Laplace transform:
\begin{eqnarray}
\sum_{m=0}^{\infty} C_{1}(m) \;  {\rm e}^{ (m+1) \, x\over t} = t \int C_1(t z) \; {\rm e}^{z \, x} \; \rm d z,
\label{lp1}
\end{eqnarray}
where $C_1(m) \equiv C_{1m}$. Using the above in eq.(\ref{pvgg}) gives 

\begin{eqnarray}
 \int C_1(t z) \; {\rm e}^{z x} \; \rm d z = 
 {2 \omega\over t} \, \left({ \omega \over 4  x}\right)^{\omega t/2} \, \Psi_v(x;0) \, {\rm e}^{-x}
\label{lp2}
\end{eqnarray}
An inverse transform then gives $C_1(tz)$. Using the relation $C_1(tz) \to C_1(m) $ then lead to the desired coefficients $C_{1m}$.  
As can be seen from the above, based on $\Psi_{v}(x; 0)$ decay for large $x$,  $C_{1m}$ can be large enough to overcome the decaying term ${\rm e}^{- {m (|x| + t \Lambda)\over t}}$ for small $m$ and $\Lambda$. This would result in a deviation of $\Psi_{v}(x; \Lambda)$ from $\Psi_{v}(x; 0)$ for finite $\Lambda$; it however may just  changes  the distribution parameters of the initial distribution without significantly affecting its shape.  This is also indicated by our numerical analysis discussed in next section.

To gain further insights in the behaviour of $\Psi_{v}(x; Y-Y_0)$, we rewrite 
eq.(\ref{pvg2}) as 

\begin{eqnarray}
\Psi_{v}(x; Y-Y_0) 
&=& \Psi_{v}(x; \infty) \; \sum_{m=0}^{\infty} {C_{1m}\over C_{10}} \;  {\rm e}^{{m (x - t \, \Lambda)\over t} }
\label{pvg2}
\end{eqnarray}
As $t >0$, the above indicates a rapid decay of the distribution for large $x$ for arbitrary $\Lambda$. For small $x$ too, the distribution stays close to $\Psi_{v}(x; \infty)$ except for the cases in which the initial distribution $\Psi_{v}(x; Y_0)$ results in very large $C_{1m}$. This behaviour  is also indicated by our numerical analysis discussed in next section for distribution of $R_1$.

In limit $N \to \infty$, we need to implement the limit  
$\omega \to \infty$. As in the purity case, the $R_1$-distribution can then be better analysed directly from $f_{v, \omega}(R_1; Y| Y_0)$,

\begin{eqnarray}
f_{v, \omega}(R_1; Y| Y_0) &=&  2^{\omega t} \;  \left(2 R_1 -t \right)^{\omega t/2}  \; {\rm e}^{-{\omega (t+1)\over t}  (t-2 R_1)} \;  \sum_{m=0}^{\infty} C_{1m} \; {\rm e}^{ - {m \omega \over t} \left((t - 2 R_1) + t \omega (Y-Y_0)\right)} 
\label{pvg3}
\end{eqnarray}
As clear from the above, $N \to \infty$ limit again leads to an abrupt transition from an  initial distribution  at $Y=Y_0$ to $f_{v}(R_1; Y-Y_0) \to \delta(t-2 R_1)$. If however, for some special initial conditions,  $C_{1m}$ values can be large enough to compensate the small values of the factors ${\rm e}^{- \, m \omega^2 \, (Y-Y_0)} $, an intermediate state for $f_v(R_1; Y)$, different from those at $Y=Y_0$ and $Y \to \infty$ may then be reached.

 Further insight in the fluctuations  of $R_1$ over the ensemble can be gained by analysing its variance, defined as $\sigma^2(R_1) = \langle \Delta R_1^2 \rangle =\langle (R_1)^2 \rangle - \langle R_1\rangle^2$ with $\langle (R_1)^n \rangle = \int  R_1^n \; f_v(R_1;Y) \;  {\rm d}R_1$); eq.(\ref{pfv}) gives  (details discussed in {\it appendix} \ref{varr1}).
\begin{eqnarray} 
 \sigma^2(R_1) = \frac{1}{a} - B e^{-4\, \Lambda}
\label{sigr1}
\end{eqnarray} 
 The above  indicates, as in purity case,  a finite width for arbitrary $\Lambda$ and finite $N$, and thereby indicating non-negligible  fluctuations of $R_1$ for many body states due to underlying complexity. The limit $N \to \infty$  case however again lead to $P_v(R_1; Y-Y_0)$ approaching a $\delta$ function even for finite $Y-Y_0$.

\section{Numerical Verification of complexity parameter based  formulation of the entropies} \label{secNumerics}

Based on the complexity parametric formulation,  different reduced matrix ensembles subjected to same global constraints, \textit{e.g.,} symmetry and conservation laws are expected to undergo similar statistical  evolution of the Schmidt eigenvalues. This in turn implies an analogous evolution  of their entanglement measures.  
Intuitively, this suggests the following: the underlying complexity of the system wipes out details of the correlations between two sub-bases, leaving their entanglement to be sensitive only to an average measure of complexity, \textit{i.e.,} $Y-Y_0$. 
Besides fundamental significance, the  complexity parameter based formulation is useful also for the following reason: for states evolving along the same path, $Y$ can  be used as a hierarchical criterion  even if they belong to different complex systems but subjected to same global constraints.

In a previous work \cite{Shekhar_2023}, we theoretically predicted that the evolutionary paths for $ \langle R_n  \rangle$ for different ensembles, with initial condition as a separable state for each of them, are almost analogous in terms of $N^2(Y-Y_0)$. The predictions were numerically verified  for three different ensembles of the state matrices $C$ with multi-parametric Gaussian ensemble density. (We recall here that the matrix element $C_{kl}$ of the state matrix $C$ corresponds to a component of a pure state $\Psi$ in product basis $|k l \rangle$ consisting of  eigenstate $|k \rangle$ of subsystem A and  $|l \rangle$ of subsystem B. A choice of variance of $C_{kl}$ changing with $l$ implies change of correlation between two subunits; this is discussed  in detail in \cite{Shekhar_2023}.  Here the ensembles considered consist of real $C$ matrices with different variance types (with $ h_{kl} = \langle C_{kl}^2\rangle - (\langle C_{kl}\rangle)^2$, $b_{kl} = \langle C_{kl}\rangle$).  The study \cite{Shekhar_2023} was however based on small-$N$ numerics as well as balanced condition $N_A=N_B=N$ and indicated only an almost collapse onto the same curve in terms of $\Lambda$. The investigation was pursued further in the  study \cite{Shekhar_2024} by a numerical comparison of the dynamics of $\langle R_1 \rangle $ and $\langle R_2 \rangle$ for large size ensembles and for unbalanced condition $N_A \not= N_B$, with $N_A= q^{L_A}$ and $q$ as the size of the local Hilbert space (i.e., the one for a  basic unit of which subsystems A and B consist of).

With focus of the present work on the distribution of entanglement entropy,  here we consider the same three ensembles to verify the  complexity parameter formulation. As discussed in section III.A of \cite{Shekhar_2024},  it is possible to choose same set of constants $Y_2, \ldots, Y_M$ for BE, PE and EE. The ensembles can briefly be described  as follows (details given in \cite{Shekhar_2023, Shekhar_2024}):

(i) {\bf Components with same variance along higher columns (BE):}   The ensemble parameters in this case are  same for all elements of $C$ matrix except those in first column: 
\begin{eqnarray}
h_{kl}  = \delta_{l1} +  \frac{ 1-\delta_{l1}}{ (1+\mu)},  \qquad b_{kl;s} = 0,  \qquad \forall \;\; k,l.
\label{vrp}
\end{eqnarray}
The substitution of the above in eq.(\ref{y}) leads to
\begin{eqnarray}
Y=-\frac{N(N_{\nu}-1)}{2M\gamma}\left[ {\rm ln} \left(1-\frac{2\gamma}{(1+\mu)} \right)\right] + c_0
\label{yrp}
\end{eqnarray}
with constant $c_0$ determined by the initial state of the ensemble. 

Choosing initial condition with $\mu \rightarrow \infty$ corresponds to  an ensemble of $C$-matrices with only first column elements as non-zero; this in turn gives $Y_0=c_0$.

(ii) {\bf Components with variance decaying as a Power law along columns (PE)}: The variance of the matrix elements $C_{kl}$ now changes, as a power law, across the column as well as row but its mean is kept zero:  
\begin{eqnarray}
h_{kl} = \frac{1}{1+\frac{k}{b}  \frac{(l-1)}{a}},\qquad  b_{kl}  =  0 \quad \forall \;\; k, l
\label{vpe}
\end{eqnarray}
where $a$ and $b$ are arbitrary parameters. Eq.(\ref{y}) then gives 

\begin{eqnarray}
Y=-\frac{1}{2M\gamma} \sum_{r_1=1}^{N} \sum_{r_2=1}^{N_{\nu}-1}  {\rm ln} \left(1-\frac{2\gamma}{1+\frac{r_1}{a}\frac{r_2}{b}}\right) + c_0
\label{ype}
\end{eqnarray}

Choosing initial condition with $b, a \rightarrow \infty$ again corresponds to  an ensemble of $C$-matrices with only first column elements as non-zero and thereby $Y_0=c_0$.

(iii) {\bf Components with variance exponentially decaying along columns (EE): } Here again the mean is kept zero for all elements but the variance changes exponentially across the column as well as row of $C$-matrix:  
\begin{eqnarray}
h_{kl} = {\rm exp}\left(-\frac{k |l-1|}{a b}\right),\qquad  b_{kl} = 0 \qquad  \forall \;\; k, l
\label{vee}
\end{eqnarray}
with $b$ as an arbitrary parameter. Eq.(\ref{y}) now gives 
\begin{eqnarray}
Y=-\frac{1 }{2M\gamma}\sum_{r_1=1}^{N} \sum_{r_2=1}^{N_{\nu}-1} \; {\rm ln} \left(1-\frac{2\gamma}{{\rm exp}(\frac{r_1}{a} \frac{r_2}{b})}\right) + c_0
\label{yee}
\end{eqnarray}
with $M=N N_{\nu}$.
Here again the initial choice of parameters $b, a \rightarrow \infty$ leads  to  a $C$-matrix ensemble  same as in above two cases and same $Y_0$.

As discussed in \cite{Shekhar_2023, Shekhar_2024}, the change of variance along the columns,  in each of the three ensembles described above, ensures a variation of  average entanglement entropy from an initial  separable state to maximally entangled state. Here the separable state corresponds to $\mu \to \infty$ for BE and  $a \to 0, b \to 0$ (equivalently $Y \to Y_0$) for PE and EE. The maximally entangled state corresponds to $\mu \to 0$ for BE and very large $a, b$ for PE and EE.

For numerical analysis of various entropies, we exactly diagonalize each matrix of the ensembles with a fixed matrix size,  and for many values of  the ensemble  parameters $a, b$ but with fixed $\gamma=1/4$.  For simplification, we consider a balanced bipartition, such that the $C$ matrix chosen for all cases is a $N \times N$ square matrix. The  obtained Schmidt eigenvalues are then used to obtain $R_1$ and $S_2$ for each matrix.  Repeating the procedure for each matrix of the ensemble, we obtain a distribution of $R_1$ and $S_2$ over the ensemble for a specific $Y$ value.
(The latter corresponds to a set of ensemble parameters for a given ensemble and can be obtained for BE, PE and EE from eq.(\ref{yrp}), eq.(\ref{ype}) and  eq.(\ref{yee}) respectively. To study its $Y$-dependence, we numerically derive  the distribution for many $Y$-values.  As in previous studies \cite{Shekhar_2023, Shekhar_2024},  here again we assume the basic units as the qubits, thus implying $q=2$.

\begin{figure}[ht!]
    \centering
    
    \includegraphics[width=\textwidth]{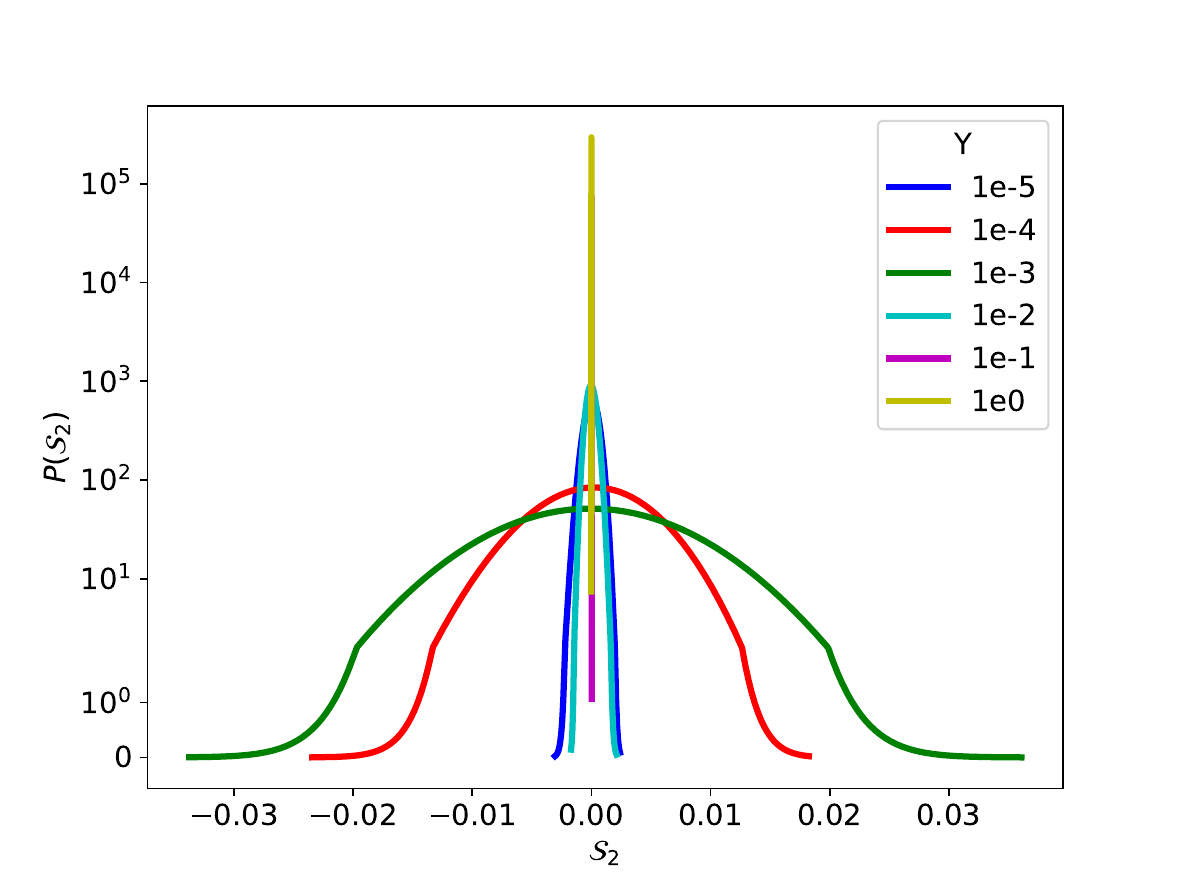}
    
    \caption{{\bf  Distribution of purity for the Brownian state ensemble  (BE, eq.(\ref{vrp})):}  The figure displays the distribution of the purity  ${\mathcal S}_2 \equiv S_2 - \langle S_2 \rangle$  for different $Y$-values, ranging from  almost separable limit to maximally entangled regimes over BE consisting of $10^5$ square $C$-matrices each of size $N=1024$. Here, to compare the distributions for different $Y$. $S_2$ is shifted by $\langle S_2 \rangle$ (the ensemble average for a fixed $Y$).  With $Y$ increasing, numerically obtained distribution changes from an initial  {\it Log-normal} form (at $Y=10^{-5}$) to {\it Beta} and finally to {\it Normal} distribution (for $Y \sim 1$). To take into account rapidly diverging distributions with increasing $Y$, a symmetric log scale is used for $y-$ axis.  The goodness of fit for each distribution has been tested by minimizing the residual sum of squares (RSS) \cite{Taskesen_distfit_is_a_2020}; details about the best fit distribution are given  in table \ref{distTableS2}. }
 \label{distS2BE}
\end{figure}

\begin{figure}[ht!]
    \centering
    
    \includegraphics[width=\textwidth]{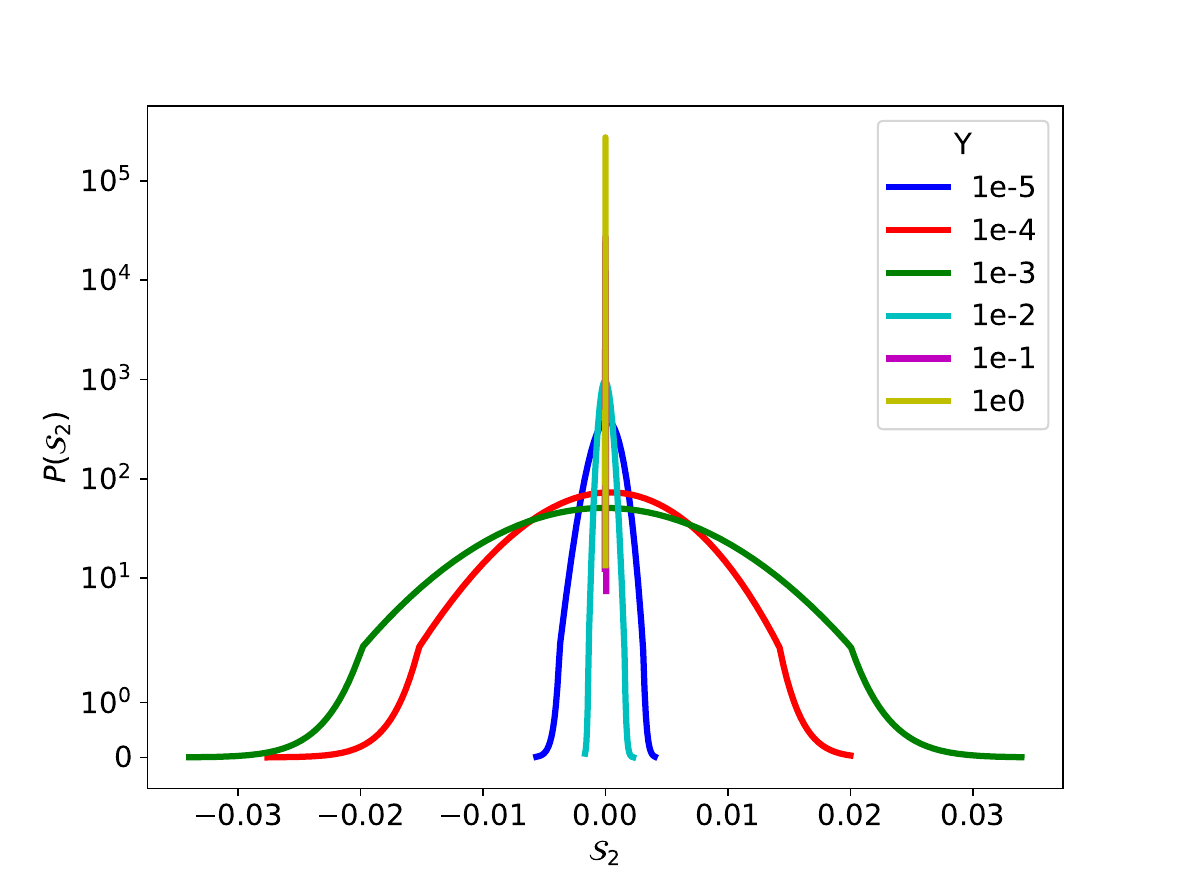}
    
    \caption{{\bf  Distribution of purity for power-law state ensemble (PE):}  The details here are same as in figure \ref{distS2BE}, except now the ensemble concerned is described by eq.(\ref{vpe}).  Here again an increasing $Y$ leads to a crossover of the numerically derived distribution from an initial  {\it Log-normal} form (at $Y=10^{-5}$) to {\it Beta} and finally to {\it Normal} distribution (for $Y \sim 1$). We note that the shape for each $Y$ is analogous to those in BE cases although best fit  parameters (table \ref{distTableS2}) are different; this is consistent our theory.}
    
    \label{distS2PE}
\end{figure}

\begin{figure}[ht!]
    \centering
    
    \includegraphics[width=\textwidth]{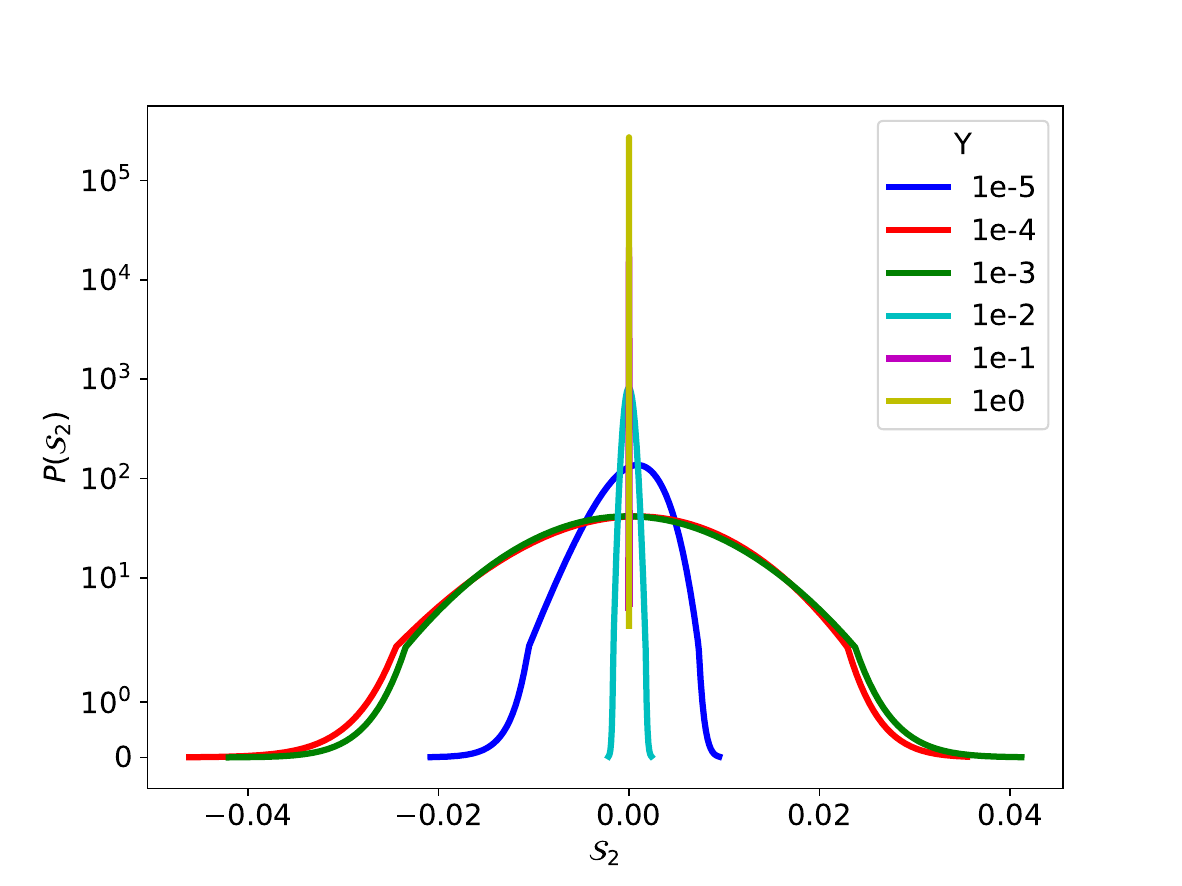}
    
    \caption{{\bf  Distribution of purity for exponential state ensemble (EE):}  The details here are same as in figures \ref{distS2BE} and \ref{distS2PE}, except now the ensemble concerned is described by eq.(\ref{vee}). In analogy with BE and PE cases,  an increasing $Y$ again leads to a change of the distribution from an initial  {\it Log-normal} form (at $Y=10^{-5}$) to {\it Beta} and finally to {\it Normal} distribution (for $Y \sim 1$). Consistent with our theory, here too the shape for each $Y$ is similar to those in BE and PE cases although best fit  parameters (table \ref{distTableS2}) vary. }
    \label{distS2EE}
\end{figure}

Figures \ref{distS2BE}-\ref{distS2EE} depict the distributions of purity  for  BE, PE and EE respectively, each case considered at six different $Y$-values. From eqs.(\ref{yrp}, \ref{ype}, \ref{yee}), a same $Y$ value for the  three ensembles requires consideration of different combinations of ensemble parameters.    As figures indicates,  the initial distribution (for $Y=Y_0 =10^{-5}$) is well-described by the  {\it Log-Gamma} behaviour ($ f(r, c) = \frac{\exp(c  r-e^r)}{\Gamma(c)}$) with $r = \frac{(S_2 -loc)}{scale}$. The crossover behaviour for small $Y-Y_0$ is well-fitted by the  {\it Log-Gamma} behaviour but the distribution parameters change (given in table \ref{distTableS2}): this is consistent with our theoretical prediction for purity case given by eqs.(\ref{ps13}, \ref{ps14}).


For $Y-Y_0 \sim 1/N$ (i.e $Y \sim 10^{-3}$ with $N=1024$) , however,  the form of the distribution changes from {\it Log-Gamma}  to  {\it Beta} distribution ($f(r, a, b) = \frac{\Gamma(a+b)}{\Gamma(a)\Gamma(b)} \,r^{a-1}(1-r)^{b-1}$).  As predicted by eq.(\ref{ps12}), the large $Y-Y_0$ limit finally converges to a narrow width {\it Gaussian} ($f(r, \sigma^2) = \frac{\exp(-r^2/2 \sigma^2 )}{\sqrt{2 \pi \sigma^2}}$).  The details of the fitted parameters i.e  $loc$, $scale$ and $c$ for each distribution are given in  table \ref{distTableS2}. As the latter indicates, the location parameter $loc$  for different $Y$ varies significantly.  For comparison of the distribution for different $Y$ values, we have shifted each distribution  by corresponding $loc$ values; this ensures same centring for them. It is worth noting from table \ref{distTableS2} that the shape of the $S_2$-distribution remains  analogous for BE, PE and EE  for a given $Y$ although their stability parameters are different. An analogy of shape for a given $Y$ verifies our theoretical prediction that the distribution of purity for the ensembles represented by eq.(\ref{rhoc}) is governed not by individual parametric details of the ensemble but by complexity parameter $Y$. Due to different stability parameters for initial $S_2$-distribution of BE, PE and EE,  a difference in their stability parameters for $Y >Y_0$ is also consistent with our formulation.

\begin{figure}[ht!]
    \centering
    
    \includegraphics[width=\textwidth]{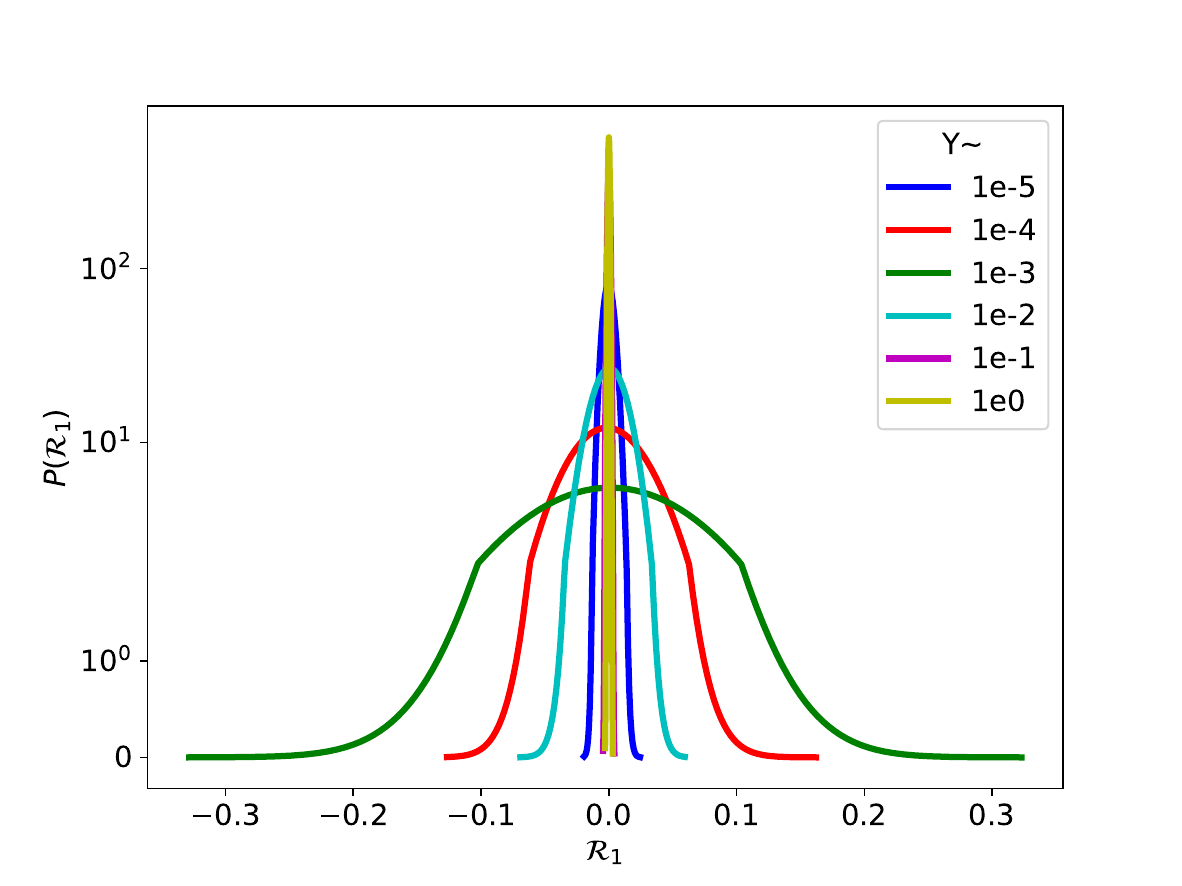}
    
    \caption{{\bf  Distribution of von Neumann entropy for Brownian state ensemble (BE, eq.(\ref{vrp})):}  The details here are same as in figure \ref{distS2BE}, except now the entanglement measure considered here is the von Neumann entropy ${\mathcal R}_1 \equiv R_1 - \langle R_1 \rangle$ with $\langle R_1 \rangle$ implying an ensemble average of $R_1$ for a fixed $Y-Y_0$.  The shape of the numerically obtained distribution  now changes from {\it Gamma} to {\it Log-Gamma} to {\it Normal} for each $Y$, with best fit parameters for each case are given in table \ref{distTableS2}. 
}
    \label{distR1BE}
\end{figure}

\begin{figure}[ht!]
    \centering
    
    \includegraphics[width=\textwidth]{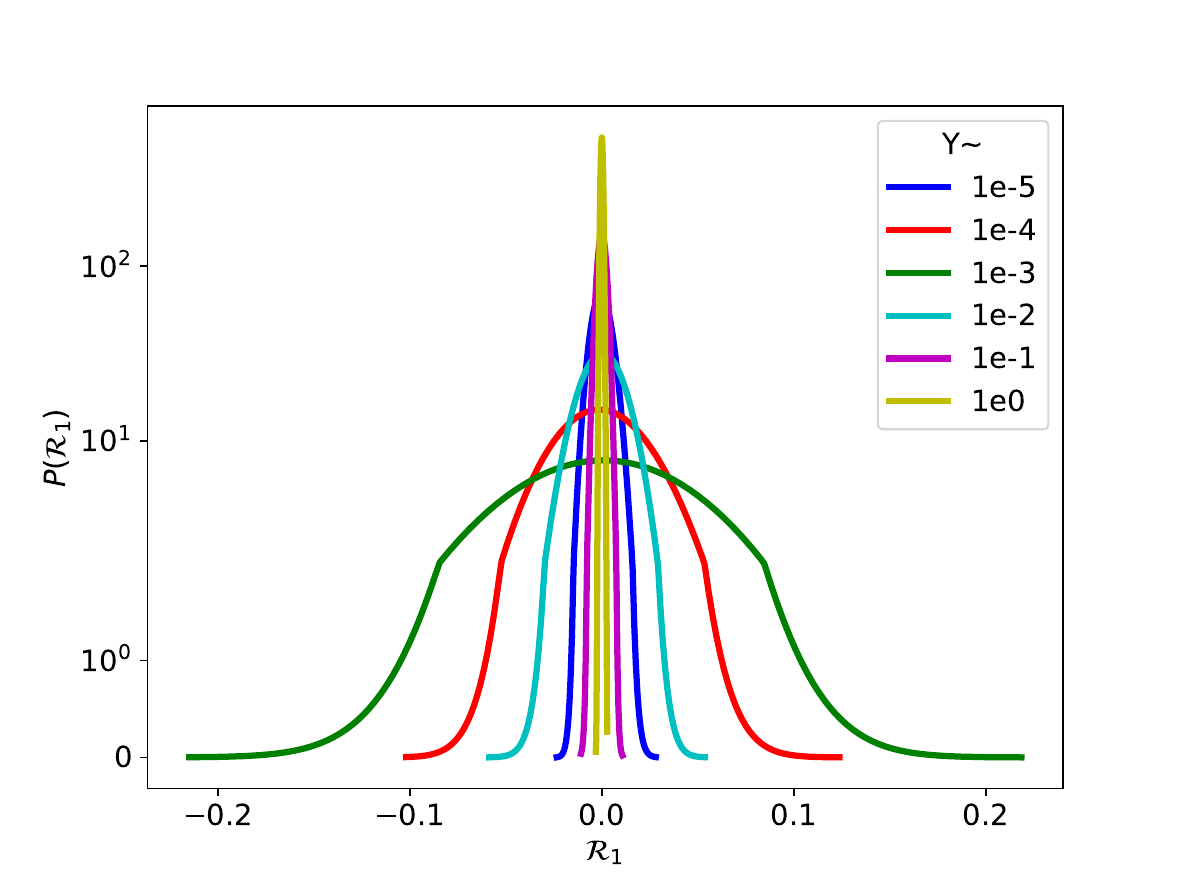}
    
    \caption{{\bf  Distribution of von Neumann entropy for  power-law state ensemble (PE, eq.(\ref{vpe})):}  The details here are same as in figure \ref{distS2PE}, except now the measure considered here is von Neumann entropy ${\mathcal R}_1 \equiv R_1 - \langle R_1 \rangle$ with $\langle R_1 \rangle$ implying an ensemble average of $R_1$ for a fixed $Y-Y_0$.
Analogous to BE case, the shape of the numerically obtained distribution changes from {\it Gamma} to {\it Log-Gamma} to {\it Normal} for each $Y$, with best fit parameters given in table \ref{distTableR1}.     
}
    \label{distR1PE}
\end{figure}

\begin{figure}[ht!]
    \centering
    
    \includegraphics[width=\textwidth]{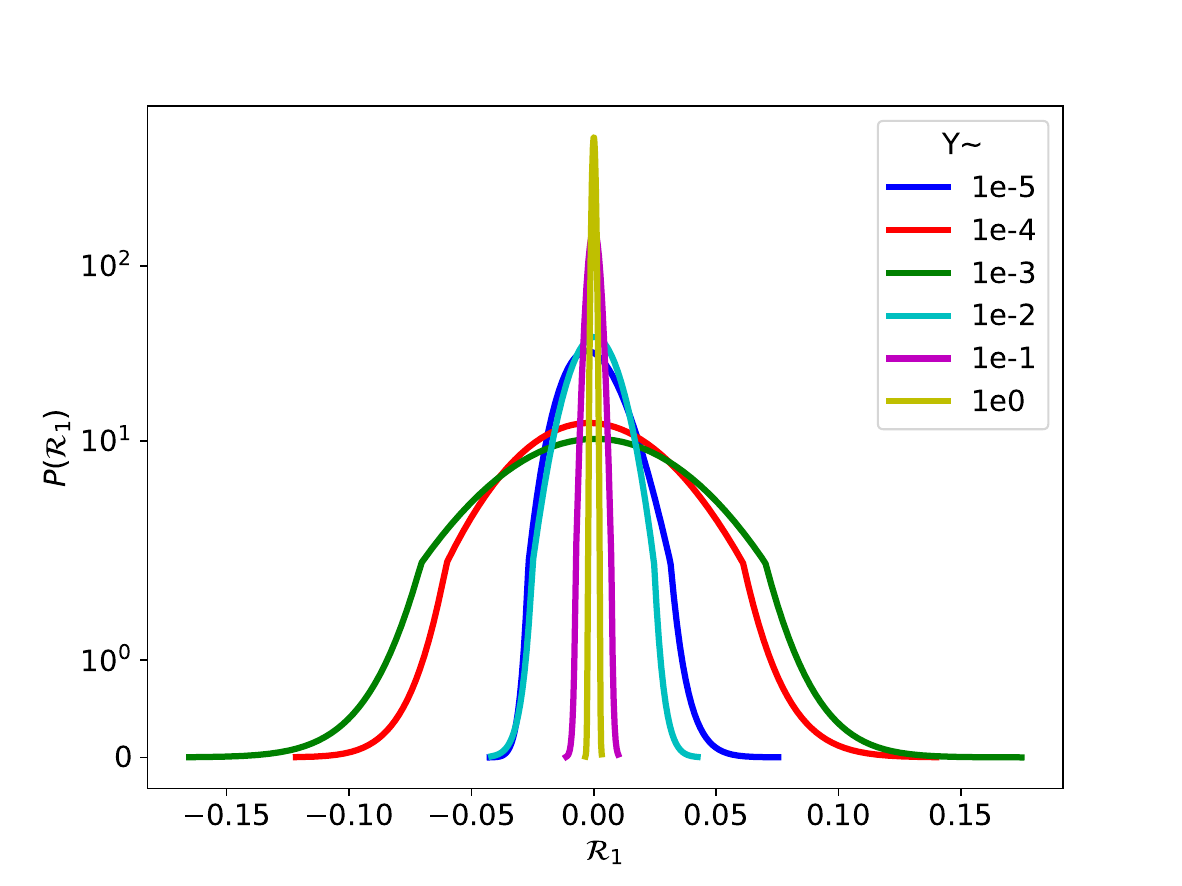}
    
    \caption{{\bf  Distribution of von Neumann entropy for exponential state ensemble (EE, eq.(\ref{vee})):}  The details here are same as in figure \ref{distS2EE}, except now the measure considered here is von Neumann entropy ${\mathcal R}_1 \equiv R_1 - \langle R_1 \rangle$ with $\langle R_1 \rangle$ implying an ensemble average of $R_1$ for a fixed $Y-Y_0$.
Similar to BE and PE cases,  an increasing $Y$ again leads to a change of the numerically derived distribution from an initial  {\it Gamma} form (at $Y=10^{-5}$) to {\it Log-Gamma} and finally to {\it Normal} distribution (for $Y \sim 1$). Consistent with our theory, here too the shape for each $Y$ is similar to those in BE and PE cases although best fit  parameters (table \ref{distTableS2}) vary.    }
    \label{distR1EE}
\end{figure}

Figures \ref{distR1BE}-\ref{distR1EE} displays the distributions of von Neumann entropies (corresponding to each purity case depicted in  Figures \ref{distS2BE}-\ref{distS2EE}).   The initial distribution at $Y_0 = 10^{-5}$ is  now well-fitted by the {\it Gamma} behaviour ($  f(r,a) = \frac{r^{a-1}\,e^{-r}}{\Gamma(a)}$) with $r = (R_1 -loc)\, / \, scale$.  As predicted by our theory,  the behaviour during crossover varies from {\it Gamma} distribution  to {\it Log-Gamma} and finally ending in a narrow Gaussian (near $Y \sim 1$); the fitted distribution parameters for each case  are given in table \ref{distTableR1}. In contrast to purity, the shape of the $R_1$-distribution changes from {\it Gamma} to {\it Log-Gamma} at $Y \sim 1/N$ (with $N=1024$).  But  as can be seen from the table \ref{distTableR1}, for a given $Y$, the shape remains  analogous for BE, PE and EE  although the distribution parameters are again different.

As clear from each of the figures, the distribution of both entropies approach a narrow distribution for small $Y$ ($\sim 10^{-5}$)  and large $Y \ge 1$, respectively.  For intermediate $Y$ values  ($Y \sim \frac{1}{N}$), however, the  distributions have non-zero finite width.  This indicates the following: as both entropies involve a sum over functions of the Schmidt eigenvalues, their local fluctuations are averaged out in separable and maximally entangled limit. For the partial entanglement region, corresponding to finite non-zero $Y$ (specifically, where average entropy changes from linear $Y$-dependence to a constant in $Y$), the eigenvalue fluctuations are too strong to be wiped out by summation, resulting in significant fluctuations of the entropies. Indeed $Y \sim \frac{1}{N}$ marks the edge between separability and fully entangled limit of the state:  states are fully separable for $Y < 1/N$ and  fully entangled  for $Y > 1/N$. As indicated in \cite{Shekhar_2023, Shekhar_2024}, many average measures of Schmidt eigenvalues undergo a  rapid change from one constant value to another in the vicinity of $Y \sim \frac{1}{N}$. 

 It is also worth noting that the distribution for $Y \sim 1$ is almost same for all three cases; this suggests that the distributions are almost  independent of ensemble details \textit{i.e.,} local constraints and reach ergodic limit as $Y \sim 1$ is reached. Further, the finite width of the distributions  also indicate that, for finite $N$, the  fluctuations  of $S_2$ and  $R_1$ around their average  can not be neglected; this has implications for ordering the states in increasing entanglement using purity or von Neumann entropy  as the criteria. We also find that the distributions are dependent on the ratio of subsystem sizes $N_A/N_B$; (the figures included here correspond to $N_A=N_B$ only).

\begin{figure}[ht!]
     \centering
    
     \includegraphics[width=\textwidth]{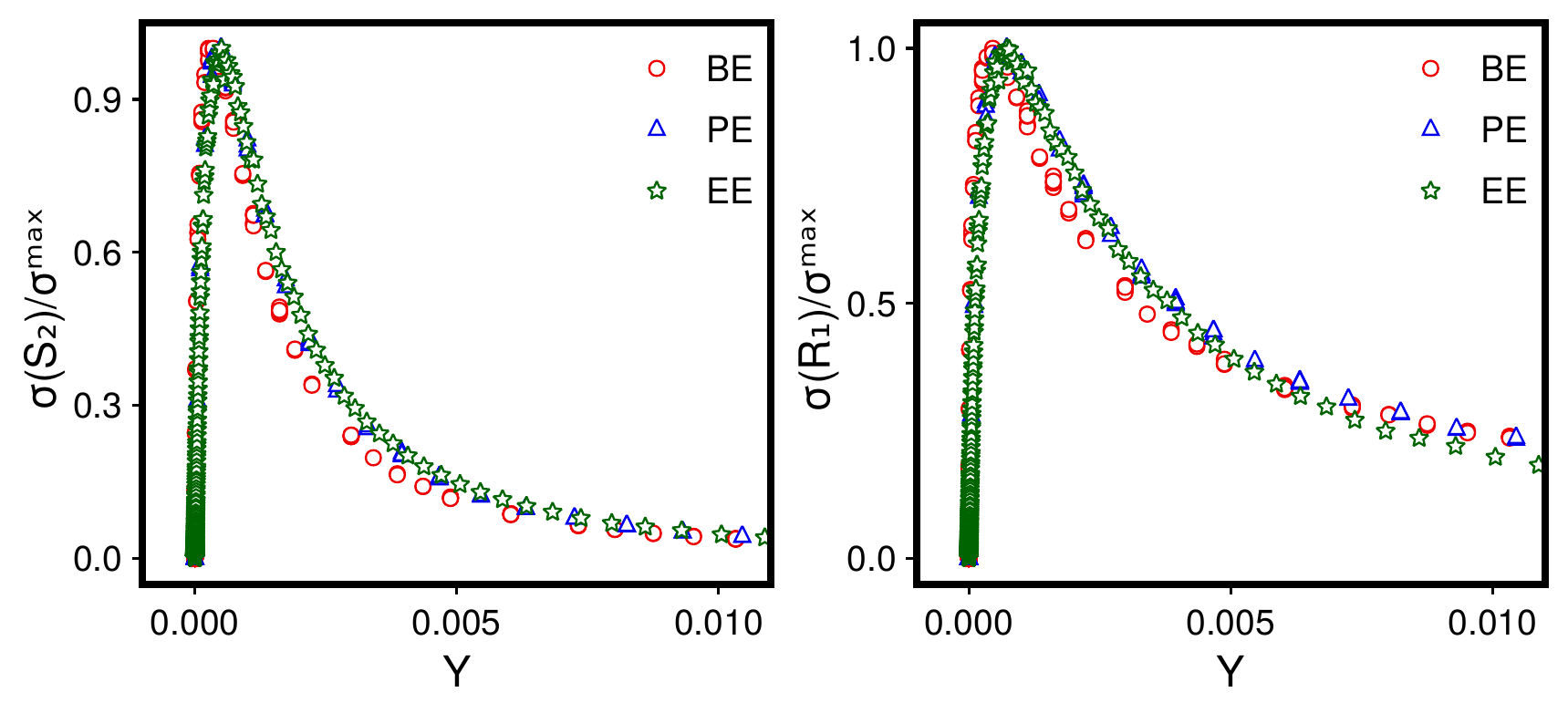}
    
     \caption{{\bf  Parametric evolution of the standard deviation of purity and entanglement entropy:}  The figure displays the evolution of the standard deviation $\sigma(S_2)$ of the purity and $\sigma(R_1)$ of von Neumann entropy (measured in units of $\sigma_{max}$) for BE, PE and EE, defined in eq.(\ref{vrp}), eq.(\ref{vpe}) and eq.(\ref{vee}),  as the complexity parameter $Y$ varies from separability to maximum entanglement limit. 
 Each ensemble used consists of $5000$ matrices of size $N=1024$. Here 
 $\sigma_{max}$ refers to the maximum standard deviation reached for an ensemble during $Y$ governed evolution. The collapse of the behaviour for different ensembles to same curve signifies the universality of the $Y$ governed evolution of the standard deviation. As can be seen from both parts, a spike  occurs at  $Y \sim 1/N$; this indicates a transition from a separable to maximum entanglement regime at  $Y \sim 1/N$. Such pattern is usually observed when the system goes under a quantum phase transition, \textit{e.g.,} thermalization to localization transition.}
    \label{stdS2R1}
 \end{figure}

Figure \ref{stdS2R1} also displays the $Y$ governed evolution of the standard deviations $\sigma(S_2), \sigma(R_1)$  of $S_2$ and $R_1$-distributions. The collapse of the curves depicting $S_2(Y)$ behaviour for BE, PE, EE for entire $Y$ range, from separability to maximum entanglement, to same curve  is consistent with our single  parametric formulation of the fluctuations of entanglement measures for pure states described by eq.(\ref{rhoc}). This is reconfirmed by a similar collapse for $R_1$.  As can also be seen from the figure, although the standard deviation of entanglement measures in the separable and maximally entangled regimes is vanishing, as $N \to \infty$, in the intermediate regime the fluctuations are orders of magnitude larger. Observation of such trends in quantum fluctuation measures are believed to indicate a quantum phase transition \cite{PhysRevLett.90.227902,PhysRevA.66.032110} and the idea has already been used to study thermalization $\to$ many-body localization transition in disordered systems \cite{Luitz2015}. 
This indeed lends credence to the state matrix models used in our analysis: although mimicking underlying complexity of the state through  Gaussian randomness, they turn out to be good enough to capture interesting many-body phenomena.

We note, while a lack of closed form formulation of the distributions (i.e eq.(\ref{ps10}) and eq.(\ref{pvg2})) for finite $N$ makes it technically challenging  to express their parameters as a function of $Y-Y_0$ and compare with our numerical results, an analogous shape for a given $Y$ and the change with increasing $Y$ is consistent with our theoretical prediction. This can also be seen from table \ref{distTableS2}. and \ref{distTableR1}: the form of the distributions for each $Y$ is analogous for BE, PE and EE although their distribution parameters are different. The latter is expected as  the initial conditions chosen in numerics for each ensemble are different. 


\begin{table}[H]
   \scriptsize
     \caption{\label{distTableS2} {\bf Details of the fitted distributions for $S_2$ in figures \ref{distS2BE}-\ref{distS2EE} :}
       The parameters for the Log-Gamma distribution $f(r, c) = \frac{\exp(c r -e^r)}{\Gamma(c)}$ (labeled as L-Gamma),  Beta distribution,  $ f(r, u, v) = \frac{\Gamma(u+v)\,r^{u-1}(1-r)^{v-1}}{\Gamma(u)\Gamma(v)}$ and  Gaussian distribution $ f(r) = \frac{\exp(-r^2/2)}{\sqrt{2 \pi}}$  with $r \equiv {(S_2-loc)\over scale}$.}


\begin{tabular}{ cp{2.2cm}p{2.2cm}p{2.5cm}p{2.2cm}p{2.2cm}p{2.5cm} }
     \hline\hline
     \multirow{2}{*}{$\mathbf{Y}$} &  \multicolumn{6}{c}{}\\
      & $10^{-5}$ & $~10^{-4}$ & $~10^{-3}$ & $~10^{-2}$ & $~10^{-1}$ & $~1$ \\ 
     \hline
     BE, $\mu$ & 100989.553 & 10013.156 & 1009.816 & 98.622 & 8.843 & 0.276 \\
     \hline
     Function & L-Gamma  & L-Gamma  & Beta  & Beta  & Beta  & Normal  \\
     LOC & 0.960 & 0.542 & 0.194 & 0.006483 & 0.00196 & 0.00195\\
     Scale & 0.0047 & 0.057 & 0.123 & 0.013 & 122.152 & 1.36e-06\\
     c  & 59.747 & 140.296 & - & - & - & - \\
     u & - & - & 26.998 & 38.163  & 129.888 & - \\
     v & - & - & 34.981 & 114.458 & 2.615e+08 & - \\
     \hline
     PE, $(a, b)$ & (1.399, 0.175) & (11.1969, 0.174) & (4.723, 4.723) & (232.862, 1.399) & (302.318, 21.869) & (859.542, 859.542)\\
     \hline
     Function & L-Gamma  & L-Gamma  & Beta  & Beta  & Beta  & Normal \\
     LOC & 0.961 & 0.659 & -34.122 & 0.0157 & 0.0044 & 0.00199 \\
     Scale & 0.0044 & 0.041 & 40.859 & 0.014 & 0.000197 & 1.454e-06\\
     c & 18.039 & 56.205 & - & - & - & - \\
     u & - & - & 3.069e+06 & 69.694 & 21.887 & - \\
     v & - & - & 578418 &  164.73 & 23.402 & - \\
     \hline
     EE, $(a, b)$ & (21.869, 0.1749) & (127.540, 0.175) & (859.542, 0.175) & (302.318, 4.723) & (859.542, 21.869) & (1020.323, 859.542)\\
     \hline
     Function & L-Gamma  & L-Gamma & Beta  & Beta  & Beta  & Normal \\
     LOC & 0.983 & 0.604 & 0.249 & 0.0313 & 0.0067 & 0.002 \\
     Scale & 0.00429 & 0.0657 & 0.169 & 298.225 & 4.331 & 1.466e-06 \\
     c & 2.34 & 48.119 & - & - & - & -\\
     u & - & - & 33.613 & 259.605 & 394.349 & -\\
     v & - & - & 43.65 & 9.727e+06 & 4.516e+06 & -\\
     \hline\hline
    \end{tabular}
  
\end{table}

\begin{table}[H]
    \scriptsize
      \caption{\label{distTableR1} {\bf Details of the fitted distributions for $R_1$ in figures \ref{distR1BE}-\ref{distR1EE} :}
        The parameters for the Log-Gamma distribution $f(x, c) = \frac{\exp(c r -e^r)}{\Gamma(c)}$ (labelled as L-Gamma), Gamma distribution  $ f(r,v) = \frac{r^{v-1}\,e^{-r}}{\Gamma(v)}$  and Gaussian distribution $ f(r) = \frac{\exp(-r^2/2)}{\sqrt{2 \pi}}$  with $r \equiv {(R_1-loc)\over scale}$.}

    \begin{tabular}{ cp{2.2cm}p{2.2cm}p{2.5cm}p{2.2cm}p{2.2cm}p{2.5cm} }
     \hline\hline
     \multirow{2}{*}{$\mathbf{Y}$} &  \multicolumn{6}{c}{}\\
      & $10^{-5}$ & $~10^{-4}$ & $~10^{-3}$ & $~10^{-2}$ & $~10^{-1}$ & $~1$ \\ 
     \hline
     BE, $\mu$ & 100989.553 & 10013.156 & 1009.816 & 98.622 & 8.843 & 0.276 \\
     \hline
     Function & Gamma & Gamma & L-Gamma  & L-Gamma & L-Gamma & Gaussian \\
     LOC & 0.093 & 0.509 & -5.122 & 8.405 & 9.218 & 9.279\\
     Scale & 0.000299 & 0.0014 & 1.702 & 0.116 & 0.01 & 0.00071\\
     c  & - & - & 564.835 & 60.837 & 89.027 & - \\
     v & 270.677 & 568.46 & - & - & - & - \\
     \hline
     PE, $(a, b)$ & (1.399, 0.175) & (11.1969, 0.174) & (4.723, 4.723) & (232.862, 1.399) & (302.318, 21.869) & (859.542, 859.542)\\
     \hline
     Function & Gamma & Gamma & L-Gamma  & L-Gamma  & L-Gamma  & Gaussian \\
     LOC & 0.0963 & 0.45 & -7.05 & 4.992 & 7.9197 & 9.26 \\
     Scale & 0.00039 & 1.634 & 1.6337 & 0.323 & 0.0759 & 0.00073\\
     c & - & - & 1018.62 & 673.32 & 861.339 & - \\
     v & 210.84 & 454.897 & - & - & - & - \\
     \hline
     EE, $(a, b)$ & (21.869, 0.1749) & (127.540, 0.175) & (859.542, 0.175) & (302.318, 4.723) & (859.542, 21.869) & (1020.323, 859.542)\\
     \hline
     Function & Gamma & Gamma & L-Gamma  & L-Gamma  & L-Gamma  & Gaussian \\
     LOC & 0.022 & -0.023 & -4.5443 & 3.056 & 7.395 & 9.255 \\
     Scale & 0.0031 & 0.0016 & 1.08677 & 0.32 & 0.032 & 0.00074 \\
     c & - & - & 796.975 & 1017.28 & 171.335 & -\\
     v & 16.628 & 370.069 & - & - & - & -\\
     \hline\hline
    \end{tabular}
   
\end{table}

\section{Conclusion} \label{secConc}

In the end, we summarize with a brief discussion of our main idea, results, and open questions. We have theoretically analysed the distribution of entanglement measures of a typical quantum state in a bipartite basis. While previous theoretical studies have mostly considered ergodic states, leading to a standard Wishart ensemble representation of the reduced density matrix, our primary focus in this work has been on the non-ergodic states, specifically, the states with their components (in bipartite basis) Gaussian distributed with arbitrary mean and variances, subjected to symmetry constraints, in addition. This in turn gives the reduced density matrix as a multi-parametric Wishart ensemble with fixed trace and permits an analysis of the entanglement  with changing system conditions, \textit{i.e.,} complexity of the system. Our approach  is based on the complexity parameter formulation  of the  joint probability distribution of the Schmidt eigenvalues which in turn leads to the evolution equations for the distributions of the  R\'enyi and von Neumann entropies. As clearly indicated by the theoretical formulations for the probability densities for finite $Y$, reconfirmed by numerical analysis of the distributions and their standard deviation,  a knowledge of the average entanglement measures is not sufficient for complex systems, their fluctuations are indeed significant,  and any hierarchical arrangement of states based solely on the averages may lead to erroneous results.

The complexity parameter based formulation of the entanglement distributions discussed here  has many potential applications. For example,  
the current quantum purification schemes for communication through noisy quantum channels are based on distillation of entanglement in the ensemble i.e by increasing the entropy for a part of the ensemble at the cost of the other part. In contrast, our formulation is based on an evolution of the entropy of a typical quantum state in the ensemble to its maximum value. 

Another potential use of our formulation is to identify various  classes of quantum states with same entanglement entropy; this follows as the entanglement entropy for each class is characterized by the complexity parameter $Y$ (besides constants of evolution). For example, consider the states belonging to two different ensembles, characterized by sets $\{h, b\}$ and $\{\tilde{h}, \tilde{b}\}$; (different ensembles here may refer e.g. different origins of the states). If however both ensembles correspond to same value of the complexity parameter as well as constant of evolution,  the distribution of their entanglement entropies over respective ensembles are predicted to be same. 

As an increasing $Y$ results in an increase in entanglement entropy, maximum entanglement for a class of quantum states can be achieved by controlling $Y$ only without considering specific details of the ensemble parameters. 
The information can also be used for a hierarchical arrangement of partially entangled states (\textit{e.g.,} revealing the flaws in their characterization based on average behaviour) as well as for phase transition studies of many body systems.

The present study still leaves many open questions. For example, a typical many-body state of a quantum system may be subjected to additional constraints and need not be represented by a multiparametric Wishart ensemble with fixed trace only.  It is then natural to query about the role of additional constraints on the  growth of entanglement measure. The question is relevant in view of the recent studies indicating different time-dependent growth rates in constrained systems \cite{PhysRevResearch.2.033020,PhysRevLett.122.250602}. A generalization of the present results to  bipartite mixed states as well as multipartite states is also very much desirable and will be presented in near future. 

Another question is regarding the explicit role played by the system parameters in the entanglement evolution. The complexity parameter depends on the system parameters through distribution parameters. A knowledge of their exact relation, however,  requires a prior knowledge of the quantum Hamiltonian, its matrix representation in the product basis as well as the nature and distribution parameters of the appropriate ensemble.  The knowledge leads to the determination of the appropriate ensemble of its eigenstates.  
A complexity parameter formulation for the Hermitian operators, \textit{e.g.,}  many-body Hamiltonians, and their eigenstates has already  been derived. As the evolution equation of the JPDF of the components of an eigenfunction can in principle be derived from that of the ensemble density of the Hamiltonian, the evolution parameter of the former must be dependent on the latter. This work in currently in progress.

\acknowledgments

We acknowledge National Super computing Mission (NSM) for providing computing resources of ‘PARAM Shakti’ at IIT Kharagpur, which is implemented by C-DAC and supported by the Ministry of Electronics and Information Technology (MeitY) and Department of Science and Technology (DST), Government of India.  One of the authors (P.S.)  is also grateful to SERB, DST, India for the financial support provided for the  research under Matrics grant scheme. D.S. is supported by the MHRD under the PMRF scheme (ID 2402341).

\newpage

\bibliographystyle{ieeetr}
\bibliography{references}

\newpage

\appendix

\section{Derivation of eqs. \eqref{ps2} and \eqref{pvn}} \label{ps2pvn}

From the definition  in eq.(\ref{psq}), we have for $k=2$,  the JPDF $P_2$ of $S_2$ and $S_1$.

For $X(\lambda)$ as an arbitrary function of eigenvalues $\lambda_1 \ldots \lambda_N$, the JPDF $P_x$ of $S_1$ and $X(\lambda)$  can be defined as 
\begin{eqnarray}
P_x(S_1,  X ; Y) 
=\int \delta_1  \; \delta_x \; P_{\lambda}(\lambda; Y) \; {\rm D}\lambda,
\label{px}
\end{eqnarray}
with $\delta_{x} \equiv \delta(X- X(\lambda) $ and $\delta_1$ defined below eq.(\ref{psq}) and $ P_{\lambda}(\lambda) \to 0$ at the two integration limits  $0 \le \lambda_n \le \infty$ $\forall n=1 \cdots N$. We note here that the constraint $\delta_1$ effectively reduces the limits to ${1\over N} \le \lambda_n \le 1$. 

A differentiation of the above equation with respect to $Y$ and subsequently using eq.(\ref{px})  gives 
\begin{eqnarray}
\frac{\partial P_x(S_1, X; Y) }{\partial Y}  
=\int \delta_1  \; \delta_x \; \frac{\partial  P_{\lambda}}{\partial Y}  \; {\rm D}\lambda  = I_1 + I_2,
\label{dpxdy}
\end{eqnarray}

with $I_1$ as,
\begin{eqnarray}
I_1 &=& - C_{hs} \; \sum_{n=1}^N \int_0^{\infty}   \;  \delta_1 \; \delta_x \;  \frac{\partial}{\partial \lambda_n}\left( \sum_{m=1}^N \frac{\beta \lambda_n}{\lambda_n- \lambda_m} + {\beta \nu} - {2 \gamma} \lambda_n \right) \;  P_{\lambda} \; {\rm D} \lambda,
\label{i1xa}
\end{eqnarray}
and,
\begin{eqnarray}
I_2 &=& C_{hs} \; \sum_{n=1}^N \int  \;  \delta_1 \; \delta_x \;  \frac{\partial^2 (\lambda_n \; P_{\lambda})}{\partial \lambda_n^2}  \;{\rm D} \lambda.
\label{i2xa}
\end{eqnarray}

Integration by parts, and, subsequently using the relation $P_{\lambda} (\lambda) \to 0$ at the integration limits,
now gives  

\begin{eqnarray}
I_1 &=&  C_{hs} \; \sum_{n=1}^N \int_0^{\infty}    \;  \left[\frac{\partial \delta_1}{\partial \lambda_n}  \; \delta_x +  \delta_1 \; \frac{\partial \delta_x  }{\partial \lambda_n}  \right] \;  \left( \sum_{m=1}^N \frac{\beta \lambda_n}{\lambda_n- \lambda_m} + {\beta \nu} - { 2\gamma} \lambda_n \right)  P_{\lambda}  \; {\rm D} \lambda. 
\end{eqnarray}

Again using $\frac{\partial \delta_a}{\partial \lambda_n}  = - \frac{\partial \delta_a}{\partial a}  \frac{\partial a(\lambda)}{\partial \lambda_n}$ with $a = S_1$ or $X$,  $I_1$ can be rewritten as 

\begin{eqnarray}
I_1
&=&   \frac{\partial   }{\partial S_1} \left(2  \; \gamma \; S_1 - {1\over 2} \beta N(N+ 2\nu-1) \right) \; P_x  +  I_0,
\label{i1xb}
\end{eqnarray}
where,
\begin{eqnarray}
I_0 = -C_{hs} \; \frac{\partial }{\partial X} \ \int    \delta_1 \;   \delta_x \; \left[ \sum_{n=1}^N \frac{\partial X }{\partial \lambda_n}  \;  \left(\sum_{m=1}^N \frac{\beta \lambda_n}{\lambda_n- \lambda_m} +  \left({\beta \nu} - { 2\gamma} \lambda_n \right) \right) \right] \; P_{\lambda} \; {\rm D} \lambda. 
\label{i0xa}
\end{eqnarray}

Similarly, a repeated partial differentiation gives

\begin{eqnarray}
I_2 
&=&  C_{hs} \;\sum_{n=1}^N \int   \;  \left[\; \delta_1 \; \frac{\partial^2  \delta_x}{\partial \lambda_n^2}   + 
2 \frac{\partial  \delta_1}{\partial \lambda_n} \; \frac{\partial \delta_x}{\partial \lambda_n} 
 + \frac{\partial^2  \delta_1}{\partial \lambda_n^2} \; \delta_x  \right] \; \lambda_n  \;   P_{\lambda} \; {\rm D} \lambda.
 \label{i2xb}
\end{eqnarray}
Again using $\frac{\partial^2 \delta_x}{\partial \lambda_n^2}  = 
\frac{\partial^2 \delta_x}{\partial x^2} \;  \left(\frac{\partial X(\lambda)}{\partial \lambda_n} \right)^2 
  -\frac{\partial \delta_x}{\partial X} \frac{\partial^2 X(\lambda)}{\partial \lambda_n^2}$
 and $\frac{\partial^2 \delta_1}{\partial \lambda_n^2}  = 
\frac{\partial^2 \delta_1}{\partial S_1^2} $,  $I_2$ can be rewritten as 
  
\begin{eqnarray}
I_2 &=&  C_{hs} \;\sum_{n=1}^N \int   \;  \left[\; \delta_1 \;\left( \frac{\partial^2  \delta_x}{\partial X^2}   
\;  \left(\frac{\partial X}{\partial \lambda_n} \right)^2 
-\frac{\partial \delta_x}{\partial X} \frac{\partial^2 X}{\partial \lambda_n^2} \right)
+ 2 \frac{\partial  \delta_1}{\partial S_1} \; \frac{\partial \delta_x}{\partial X} \frac{\partial X}{\partial \lambda_n} 
+ \frac{\partial^2  \delta_1}{\partial S_1^2} \;  \delta_x  \right]  \; \lambda_n \; P_{\lambda} \; {\rm D} \lambda. \nonumber \\
\label{i2xc}
\end{eqnarray}

Based on the details of function $X(\lambda)$, the integrals $I_1$ and $I_2$ can further be reduced.  Here we give the results for $S_2$ and $R_1$.

{\bf Case $X=S_2$:} Proceeding similarly for $X(\lambda)=  \sum_n \lambda_n^2$, we have, 
from eq. (\ref{i1xb}) and eq. (\ref{i0xa}),

\begin{eqnarray}
I_1 &=&  \frac{\partial   }{\partial S_1} (2  \; \gamma \; S_1-{1\over 2} \beta N(N + 2\nu-1)) \; P_2  + I_0.
\label{i1s2a}
\end{eqnarray}
and

\begin{eqnarray}
I_0 &=&   -2 C_{hs} \; \frac{\partial   }{\partial S_2} \int    \delta_1 \;   \delta_2 \; \left( \sum_{m, n=1}^N \frac{\beta \lambda_n^2}{\lambda_n- \lambda_m} + {\beta \nu} \sum_{n=1}^N \;\lambda_n - { 2\gamma} \sum_{n=1}^N \;\lambda_n^2  \right)  P_{\lambda} \; {\rm D} \lambda. 
\end{eqnarray}
Using the equality $2 \; \sum_{m, n=1}^N \frac{\lambda_n^2}{\lambda_n- \lambda_m} = 
\sum_{m, n=1}^N \frac{\lambda_n^2-\lambda_m^2}{\lambda_n- \lambda_m} = 2 (N-1) \sum_n \lambda_n$,  $I_0$ can be rewritten as
\begin{eqnarray}
I_0 &=&  -C_{hs} \; \frac{\partial   }{\partial S_2}  \int    \delta_1 \;   \delta_2 \;  \left( 2( N-1 +\nu) \beta \; S_1 - { 4\gamma} \; S_2 \right)  P_{\lambda} \; {\rm D} \lambda \\
 &=&  - \frac{\partial   }{\partial S_2}    \left( 2( N-1 +\nu) \beta \; S_1 - { 4\gamma} \; S_2 \right)  P_2. 
\label{i0s2a}
\end{eqnarray}
Proceeding similarly, from eq. \eqref{i2xc} we have 
\begin{eqnarray}
I_2 &=&  4 \frac{\partial^2}{\partial S_2^2} I_3 - 2 \frac{\partial }{\partial S_2}(S_1 P_2) + 4 \frac{\partial^2}{\partial S_1 \partial S_2}(S_2 P_2) + \frac{\partial^2}{\partial S_1}(S_1 P_2).
\label{i2s2a}
\end{eqnarray}

Substitution of eq.(\ref{i1s2a}) and eq.(\ref{i2s2a}) in eq.(\ref{dpxdy}) for $X=S_2$ now leads to

\begin{eqnarray}
\frac{\partial P_2}{\partial Y}= 4 \frac{\partial^2 (S_2 \; P_2)}{\partial S_2 \partial S_1} + 4 \frac{\partial^2 I_3}{\partial S_2^2} + \frac{\partial^2 (S_1 \; P_2)}{\partial S_1^2} + 2 \frac{\partial (Q_2 \; P_2)}{\partial S_2} +\frac{\partial (Q_1 \; P_2)}{\partial S_1},
\label{ps2app}
\end{eqnarray}

with $Q_2 \equiv (2 \gamma S_2 - \beta(N + \nu - 1)S_1 - S_1)$, $Q_1= 2 \gamma  S_1-{\beta\over 2} N (N + 2 \nu - 1)$ and 

\begin{eqnarray}
I_3 = \int \prod_{k=1}^2 \; \delta_{S_k} \; \left(\sum_n \lambda_n^3 \right) \;  P_{\lambda} \; {\rm D}\lambda = \int \; S_3 \; P_3(S_1, S_2,S_3) \; {\rm d}S_3.
\label{i3app}
\end{eqnarray} 

{\bf Case $X=R_1$:} Following from above, we have for $X(\lambda) =-\sum_n \lambda_n \log \lambda_n$, 

\begin{eqnarray}
I_1 &=&   \frac{\partial   }{\partial S_1} (2  \; \gamma \; S_1-{1\over 2} \beta N(N+2\nu-1)) \; P_v(S_1, R_1, Y) + I_0,
 \label{i1r1a}
\end{eqnarray}
and
\begin{eqnarray}
I_0  &=& C_{hs} \frac{\partial}{\partial R_1}\; \int  \; \delta_1 \; \delta_v \left[ \sum_{m,n=1}^N \frac{\beta \lambda_n \left(1+\log \lambda_n \right)}{\lambda_n- \lambda_m} + {\beta \nu} \sum_n \left(1+\log \lambda_n \right) - { 2\gamma } \sum_n \lambda_n \left(1+\log \lambda_n \right) \right] \; P_{\lambda}  {\rm D} \lambda.  \nonumber \\
\label{i0r1a}
\end{eqnarray}

Further, using the  relation $\sum _{n \neq m} \frac{\lambda_n \log \lambda_n}{\lambda_n - \lambda_m} 
\approx \frac{N (N-1)}{2} - \frac{(N-1)}{2} R_0$ where $R_0(\lambda) \equiv -\sum_n \log \lambda_n $ (derived in {\it appendix J} of \cite{Shekhar_2023}), 
eq.(\ref{i0r1a}) can be rewritten as 
\begin{eqnarray}
    I_0 = \frac{\partial}{\partial R_1}\left[\left(\beta N (N + \nu - 1) + 2 \gamma (R_1 - S_1) - \beta \frac{N_{\nu}}{2} R_0\right) P_v\right].
\label{i0r1b}
\end{eqnarray}

Substitution of eq.(\ref{i1r1a}) and eq.(\ref{i0r1b}) in eq.(\ref{i1xb}),  we have for $X= R_1$,

\begin{eqnarray}
    I_1 &=& \frac{\partial }{\partial S_1} \left[\left(2  \; \gamma \; S_1-{1\over 2} \beta N(N+2\nu-1)\right)\; P_v \right]\nonumber \\
    &+& \frac{\partial}{\partial R_1}\left[\left(\beta N (N + \nu - 1) + 2 \gamma (R_1 - S_1) - \beta \frac{N_{\nu}}{2} R_0\right) P_v\right]. \label{i2ci}
\end{eqnarray}

Substituting $X=-\sum_n \lambda_n \log \lambda_n$ in eq.(\ref{i2xc}), $I_2$ can be written as 

\begin{eqnarray}
    I_2 &=& \frac{\partial^2}{\partial R_1^2}\left((S_1 - 2 R_1)P_v + J_1\right) - 2 \frac{\partial^2}{\partial S_1 \partial R_1}\left((S_1 - R_1)P_v\right) \nonumber \\
    &+& N \frac{\partial}{\partial R_1}P_v + \frac{\partial^2}{\partial S_1^2}(S_1 P_v), \label{i1c}
\end{eqnarray}
with 
\begin{eqnarray}
J_{k}  &=& \int  \delta_{R_1} \; \delta_{S_1} \; \left[\sum_n (\lambda_n)^k (\log \lambda_n)^{k+1} \right] \; P_{\lambda} \; {\rm D}\lambda, \label{jk}
\end{eqnarray}

where $\delta_{R_1} \equiv \delta(R_1+\sum_n \lambda_n \; {\rm log} \lambda_n)$ and  $\delta_{S_1} \equiv \delta(S_1-\sum_n \lambda_n)$. \\
$\Rightarrow$
\begin{eqnarray}
    \frac{\partial P_v}{\partial Y} &=&  2\frac{\partial^2 }{\partial R_1 \partial S_1} \left[(R_1-S_1) \; P_v\right]+  \frac{\partial^2 }{\partial R_1^2} \left[(S_1-2 R_1)P_v +J_1\right]+ \frac{\partial^2 }{\partial S_1^2} (S_1 \; P_v) \nonumber \\
    &+& \frac{\partial}{\partial R_1}\left[\left(\beta N (N + \nu - 1) + 2 \gamma (R_1 - S_1) - \beta \frac{N_{\nu}}{2} \langle R_0 \rangle + N\right) P_v\right] \nonumber \\
    &+& \frac{\partial }{\partial S_1} \left[\left(2  \; \gamma \; S_1-{1\over 2} \beta NN_{\nu}\right)\; P_v \right].
    \label{pvnapp}
\end{eqnarray}

To express the above integral in terms of $P_v$, we can  approximate them as follows. We rewrite $J_k$ as 
\begin{eqnarray}
J_k =\int  T_k \; P(R_1,  S_1, T_k) \; {\rm d}T_{k},
\label{jk0}
\end{eqnarray}
with 
 \begin{eqnarray}
 P(R_1,  S_1,  T_k) =\int \delta_{R_1} \; \delta_{S_1}  \delta_{T_k}
 \;  P_{\lambda} \; {\rm D}\lambda,
\label{pvt}
\end{eqnarray}
and $T_k= \sum_n (\lambda_n)^{k} (\log \lambda_n)^{k+1}$. Using condition probability relation $ P(R_1, S_1,  T_k) =  P(T_k | R_1,S_1) \, P(R_1, S_1)$ in eq.(\ref{jk0}), we have 
\begin{eqnarray}
J_k &=&  P(R_1, S_1) \, \int  T_k \; P(T_k | R_1,S_1)  \; {\rm d}T_{k}, \\
&=& P(R_1, S_1) \, \langle T_k \rangle_{R_1, S_1} 
\label{jk01}
\end{eqnarray}
We note that  $T_k$ varies very slowly with $R_1$ and $S_1$; this permits the approximation $\langle T_k \rangle_{R_1, S_1} \approx \langle T_k \rangle$.

\vspace{0.5in}

\section{Solution of eq. (\ref{pf})} \label{solns2}

For large $Y$, $\langle S_3 \rangle$ approaches  a constant: $\langle S_3 \rangle \sim {1\over N^2}$. Again using separation of variables and for arbitrary initial condition at $Y=Y_0$, the solution can be given as
\begin{eqnarray}
\Psi(x; Y) = {\rm e}^{- E \; (Y-Y_0)} \; V(x; E),
\label{psa1}
\end{eqnarray}

where $V(x, E)$ is the solution of the differential equations
\begin{eqnarray}
\frac{{\rm d}^2 V}{{\rm d} x^2}  + 2 x  \; \frac{{\rm d} V}{{\rm d} x} + {d\over 2 \omega}  \; V =0,
\label{psa2}
\end{eqnarray}
with
$\eta \approx  4 \, \omega + 2 \gamma \approx 4 \omega$, $d=d_0+E$ with $E$ as an arbitrary semi-positive constant and $d_0=(\gamma-\beta N N_{\nu}) \, {\omega\over 2} + \omega^2 + 4 \gamma +1$.
In large $N$ limit, $d_0$ can further be approximated as $d_0 \approx  {\omega\over 2} \, (2\omega - \beta N N_{\nu}) > 0$ (assuming $\omega \gg N N_{\nu}, \gamma$). 
A solution of eq.(\ref{psa2}) can now be given as 

\begin{eqnarray}
V(x; E) = {\rm e}^{-{x^2}}  \; \left(  c_{1}(E) \; H_{2\mu} \left( x \right) + c_2(E) \; _1F_1 \left(-{\mu}, {1\over 2},  x^2\right)\right),
\label{psa3}
\end{eqnarray}

with $\mu = \frac{d - 4 w}{8 w}\approx {d\over 8 \omega}$. The constants of integration $c_1, c_2$  are  determined from the boundary conditions. Here $H_{\mu}(x)$ is the  Hermite's function and 
$_1F_1(-\mu,\zeta; z) $ is the confluent Hypergeometric function
defined as $_1F_1(-\mu,\zeta; z) =\sum_{k=0}^{\infty} {(-\mu)_k \, z^k\over k! \, \zeta_k }$ with symbol $a_k \equiv  \prod_{n=0}^{k-1} (a+n)$.

Further, noting that $$ H_{2\mu} \left(x \right) ={2^{2\mu} \sqrt{\pi} \over \Gamma{({1-2\mu\over 2})}}  \; _1F_1 \left(-{\mu}, {1\over 2}, x^2 \right) - {2^{2\mu+1} \sqrt{\pi}  \over \Gamma{(-{\mu})}} \; x \; _1F_1 \left({1-2\mu\over 2}, {3\over 2}, x^2 \right), $$
eq.(\ref{psa3}) can be rewritten as  

\begin{eqnarray}
V(S_2; E) =  {\rm e}^{-{x^2} } \; \left( c_3(E)  \, _1F_1 \left(-{\mu}, {1\over 2}, x^2 \right)  + c_4(E) \, x \,  _1F_1 \left({1-2\mu\over 2}, {3\over 2}, x^2 \right)\right), 
\label{psa4}
\end{eqnarray}
with $c_3(E)={2^{2\mu} \sqrt{\pi} \over \Gamma{({1-2\mu\over 2})}} c_1 +c_2$ and  $c_4(E) = -  {2^{2\mu+1} \sqrt{\pi} \; c_1  \over \Gamma{(-{\mu})}}$. 

As $E$ is an arbitrary constant and $g(S_1)$ is an arbitrary function of $S_1$, it is easier to write, without loss of generality, $E=|d_0| \,m = 8 \omega \mu_0 m$ with $m$ as a semi-positive integer, satisfying the condition $E \ge 0$.  This gives $d=|d_0|(m+s_0))$ and thereby $\mu \equiv \mu_m ={|d_0| \over 8 \, \omega}  (m + s_0)$ with $s_0= 1, 0,-1$ if $d_0 >0,  0, < 0$ respectively. 
The general solution for $\Psi(x,Y)$ can  now be written as 

\begin{eqnarray}
\Psi(x; Y)=   {\rm e}^{-{x^2}} \sum_{m=0}^{\infty} \; {\rm e}^{- 8 \omega \mu_0 \, m \; (Y-Y_0)} \, \left( c_{3m}  \,  _1F_1\left(-{\mu_m}, {1\over 2}, x^2 \right)  + c_{4m} \, x\,  _1F_1 \left({1-2\mu_m \over 2}, {3\over 2}, x^2 \right) \right). \nonumber \\
\label{psa5}
\end{eqnarray}

Further, following the relation $$ x^2 \, _1F_1\left({1-\mu_m \over 2}, {3\over 2}, x^2 \right)= {1\over 2} \; \left[_1F_1 \left({1-\mu_m \over 2}, {1\over 2}, x^2 \right) - _1F_1 \left(-{1+\mu_m \over 2}, {1\over 2}, x^2 \right)   \right],$$we have, for large $\mu_m$, $x^2 \, _1F_1 \left({1-\mu_m \over 2}, {3\over 2}, x^2 \right) \approx 0$. 
For $x >0$, the contribution from the second term in eq.(\ref{psa5}) can then be neglected, leading to following form

\begin{eqnarray}
\Psi(x; Y) \approx   {\rm e}^{-{x^2} } \, \sum_{m=0}^{\infty} \; {\rm e}^{- 8 \, \omega \, \mu_0 \, m \; (Y-Y_0)}  \,  _1F_1 \left(-{\mu_m}, {1\over 2}, x^2 \right)  \, C_{m},
\label{psb5}
\end{eqnarray}
with $C_m$ as constants to be determined from initial conditions. 


To proceed further, we use following relation: for $a \to \infty$,  $|ph(a)| \le \pi-\delta$ and $b$, $z$ fixed

\begin{eqnarray}
F\left(-a,b,z\right)\sim\left(z/a\right)^{(1-b)\over 2}\frac{e^{z/2}
\Gamma\left(1+a\right)}{\Gamma\left(a+b\right)}\*\left(J_{b-1}\left(2\sqrt{az}
\right)\sum_{s=0}^{\infty}\frac{p_{s}(z)}{(-a)^{s}}-\sqrt{z/a} \; J_{b}\left(2\sqrt{
az}\right)\sum_{s=0}^{\infty}\frac{q_{s}(z)}{(-a)^{s}}\right),\nonumber \\
\label{fap} 
\end{eqnarray}

with $p_{k}(z)=\sum_{s=0}^{k}\genfrac{(}{)}{0.0pt}{}{k}{s}{\left(1-b\right)_{k-s}}
z^{s}c_{k+s}(z),$ $q_{k}(z)=\sum_{s=0}^{k}\genfrac{(}{)}{0.0pt}{}{k}{s}{\left(2-b\right)_{k-s}}
z^{s}c_{k+s+1}(z),$ where $(k+1)c_{k+1}(z)+\sum_{s=0}^{k}\left(\frac{bB_{s+1}}{(s+1)!}+\frac{z(s+1)B_{s+2
}}{(s+2)!}\right)c_{k-s}(z)=0.$ For large $\mu_m$ values, substitution of eq.(\ref{fap}) in eq(\ref{psb5}) now leads to 


\begin{eqnarray}
\Psi(x; Y-Y_0) &\approx & {\rm e}^{-{x^2\over 2} }  \,  \sum_{m=0}^{\infty} \; C_{m}  \;    \left({x^2\over \mu_m}\right)^{1\over 4} \;  {\mathcal J}_m \;  {\rm e}^{- 8 \, \omega \, \mu_0 \, m \; (Y-Y_0)},
\label{ps9a}
\end{eqnarray}
with,
\begin{eqnarray}
{\mathcal J}_m = J_{-1/2}\left(2\sqrt{\mu_m x^2}
\right)\sum_{s=0}^{\infty}\frac{p_{s}(x^2)}{(\mu_m)^{s}}-\sqrt{x^2 \over \mu_m} \; J_{1/2}\left(2\sqrt{\mu_m \, x^2}\right)\sum_{s=0}^{\infty}\frac{q_{s}(x^2)}{(\mu_m)^{s}}.
\label{jj}
\end{eqnarray}

As $\mu_m$ is large for  $m$ large (eq.(\ref{mum})),  eq.(\ref{jj}) can further be simplified  by approximations  $\sum_{s=0}^{\infty}\frac{v_{s}(x^2)}{(\mu_m)^{s}} \approx \frac{v_{0}(x^2)}{(\mu_m)}$ with $v_s \equiv p_s$ or $q_s$. This along with the relations $J_{-1/2}(x) =\sqrt{{2 \over \pi x}} \, \cos x$ and $J_{1/2}(x) =\sqrt{{2 \over \pi x}} \, \sin x$ 
gives 

\begin{eqnarray}
{\mathcal J}_m 
&\approx &  {( \pi^2 \, \mu_m \, x^2)^{-1/4}} \; \cos\left(2 \sqrt{x^2 \mu_m}\right), 
\label{jj1}
\end{eqnarray}
where the term $\sqrt{\frac{x^2}{\mu_m}} \sin\left(2 \sqrt{x^2 \mu_m}\right)\left(\frac{3 - x^2}{12}\right)$ is neglected.

\section{Solution of eq.(\ref{pfv})} \label{solnr1}

Using separation of variables again, a solution of eq.(\ref{pfv}) for arbitrary initial condition at $Y=Y_0$ can be written as 

\begin{eqnarray}
f_{v, \omega}(R_1; Y) = {\rm e}^{-E \; (Y-Y_0)} \; V(R_1; E),
\label{pra0}
\end{eqnarray}
with $V(R_1, E)$  as the solution of the differential equation 

\begin{eqnarray}
(t-2 R_1) \; \frac{{\rm d}^2 V}{{\rm d} R_1^2}  +(a R_1 + b) \; \frac{{\rm d} V}{{\rm d} R_1} + d \; V  =0, 
\label{pra1}
\end{eqnarray}
with $a=2 \gamma +2 \, \omega \approx 2 \omega$ (assuming $\gamma$ finite),  
$b= \beta  \, (N_{\nu} - \nu) \, N - {\beta \over 2}   N_{\nu} \, \langle R_0 \rangle + N + 2 \, \omega - 2\gamma -4 $ where $R_0=-\sum \log \lambda_n$, $d =E+d_0$ and  $d_0 ={\beta\over 2} \, \omega \, N \, N_{\nu}  + \omega^2 + (2-2\gamma) \; \omega - 4$, $t=1+\langle T_1 \rangle \sim 1+\langle R_1 \rangle^2$, $N_{\nu} = N+2\nu-1$ and $E$ as an arbitrary positive constant determined  by the initial conditions on $V$. 
A solution of eq.(\ref{pra1}) can now be given as
\begin{eqnarray}
V(R_1; E) =  \left({8 x\over a}\right)^{\alpha} \; \left( c_1 \; U \left[\alpha+{ d\over a}, \alpha+1, x \right] + c_2 \; L \left[-\left(\alpha + {d\over a}\right), \alpha, x\right] \right), \nonumber \\
\label{solpv1}
\end{eqnarray}

with  $x \equiv -a (t-2 R_1)/4$ (used for compact writing), $ c_1, c_2$ as unknown constants of integration, $\alpha={1\over 4}(a t+2 b +4) $. Here
$U$ is a confluent Hypergeometric function of first kind,  defined as 
$$U(a,b,x)={\Gamma(b-1)\over \Gamma(a)} \, x^{1-b} \, _1F_1(a-b+1, 2-b, x) + {\Gamma(1-b)\over \Gamma(a-b+1)}  \, _1F_1(a, b, x), $$ 
with $_1{\tilde F}_1$ as the Hypergeometric function defined as follows: $ _1{F}_1(a,b,x) = \sum_{k=0}^{\infty}  {(a)_k \, x^k \over (b)_k \, k!}$.
The above definition  is valid for $b$  not an integer.
Here, for $b$ as an integer,  it should be replaced by $b_1$ where $b$ is integer limit of $b_1$.
Further $L_a^b(x) \equiv L(a, b;x)$ is the associated Laguerre polynomial, defined as 
$$L(a,b,x)={\Gamma(a+b+1)\over \Gamma(a+1) \, \Gamma(b+1)} \, _1{ F}_1(-a, b+1, x).$$

Using the above relations, eq.(\ref{solpv1}) can be rewritten as 
\begin{eqnarray}
V(R_1; E) =  \left({8 x\over a}\right)^{\alpha} \; \left( C_1  \; _1F_1 \left[\alpha + {d\over a}, \alpha+1, x\right]
+ C_2 \, x^{-\alpha} \,  _1F_1\left[{d\over a}, 1-\alpha, x \right] 
 \right).
\label{solpv2}
\end{eqnarray}
Here again, with  both $E$ and $\omega$ arbitrary in eq.(\ref{pra0}),   we can write, without loss of generality, $E= m \, |d_0| $ where  $m$ is an arbitrary non-negative integer.

In large $N$ and $\omega$ limits,  we have $a \approx 2 \omega$, $b \approx \beta N N_{\nu} + 2 \omega - {\beta \over 2} N_{\nu} \langle R_0 \rangle $ with $\langle R_0 \rangle \sim N \log N$ \cite{Shekhar_2023} , $d_0 \approx \omega \, \left({\beta \over 2} N N_{\nu} +\omega\right) >0$. The general solution can now be written as 

\begin{eqnarray}
f_{v, \omega}(R_1; Y)=\sum_{m=0}^{\infty} {\rm e}^{- \, m \, |d_0|   \; (Y-Y_0)} \; V_m(R_1; E),
\label{pra4}
\end{eqnarray}
with,
\begin{eqnarray}
V_m(R_1; E) &=& \left({ 4 x\over \omega}\right)^{\alpha}  \; \left( C_{1m} \; _1F_1\left(\alpha +{ d_m\over 2\omega},    \alpha+1, x \right) + C_{2m} \; x^{-\alpha} \, _1F_1\left({d_m \over 2 \omega}, 1-\alpha, x \right) \right), \nonumber 
 \\
 \label{vm1}
\end{eqnarray}

with  $x \equiv -{\omega\over 2}  (t-2 R_1) \sim  -{\omega\over 2}  (1 + \langle R_1 \rangle^2-2 R_1) $ and $d_m = d_0(1+m) \approx  (m+1) \, \omega \, \left({\beta \over 2} N N_{\nu} +\omega\right)$ and $\alpha  \approx {1\over 2} (\omega \, t+b) $. Further, as the  contribution from the second term in eq.(\ref{vm1}) is negligible as compared to first term, and we can approximate $V_m(R_1; E) = \left({ 4x\over \omega}\right)^{\alpha}  \; C_{1m} \; _1F_1\left( \alpha +{d_m,\over 2\omega},  \alpha+1,  x \right) $
\begin{eqnarray}
 f_{v, \omega}(R_1; Y)=   \left({ 4x\over \omega}\right)^{\alpha}  \; {\rm e}^x \; \sum_{m=0}^{\infty} C_{m} \; {\rm e}^{- \, {m  \, \mid d_0\mid  \; (Y-Y_0)}} \;  \; _1F_1\left(1-{d_m,\over 2\omega},  \alpha+1,  - x \right). 
\label{pra5}
\end{eqnarray}

\section{Calculation of variance of purity from eq.(\ref{pf})} \label{vars2}

Eq.(\ref{pf}) describes the $Y$-evolution of the distribution $f_2$ of the purity $S_2$.  This can further be used to derive the $Y$-evolution of  the variance of $S_2$. 
Using the definition of variance $\langle \Delta S_2^2 \rangle \equiv \langle S_2^2 \rangle - \langle S_2 \rangle^2$ and the $n^{th}$ order moment  as $  \langle S^n_2(Y) \rangle = \int S^n_2 \,f_2(S_2, Y)\,dS_2$ with $\int f_2 \, dS_2=1$, we have
\begin{eqnarray}
 \frac{\partial \langle \Delta S_2^2 \rangle}{\partial Y} &=& \frac{\partial \langle S_2^2 \rangle}{\partial Y} - 2 \langle S_2 \rangle \frac{\partial \langle S_2 \rangle}{\partial Y},
 \label{pvars2}
\end{eqnarray}

Substituting the definition of $\langle \Delta S_2^2 \rangle$, along with
eq.(\ref{pf}), in the above equation, with
$a = 4\langle S_3 \rangle, \, \eta \approx 4 \omega, \, b = -(N_a+N_b-1) \beta$, and repeated partial integration  of the terms in right side of the above equation gives

\begin{eqnarray}
\frac{\partial \langle S_2 \rangle}{\partial Y} &=& -b - \eta \,\langle S_2 \rangle \label{s2n1} \\
\frac{\partial \langle S_2^2 \rangle}{\partial Y} &=&  2a - 2 \eta \, \langle S_2^2 \rangle - 2 \, b \, \langle S_2 \rangle \label{s2n2}
\end{eqnarray}

Substitution of eq. (\ref{s2n1}) and eq.(\ref{s2n2}) in eq.(\ref{pvars2}) then leads to
\begin{eqnarray}
    \frac{\partial \langle \Delta S_2^2 \rangle}{\partial Y} &=& 2 \, a - 2 \eta \langle \Delta S_2^2 \rangle.
 \end{eqnarray}
The above equation can now be solved to give
\begin{eqnarray}
    \langle \Delta S_2^2 \rangle &=& \frac{a}{\eta} - A \, e^{-2 \eta Y},
\end{eqnarray}
where the constant of integration $A$ is determined by the initial condition which is system dependent. Nonetheless, since, $\omega \sim N^2$, and $\langle S_3 \rangle \sim \frac{1}{N^2},$we infer that in the stationary limit, $Y \to \infty$, $\langle \Delta S_2^2 \rangle \sim \frac{1}{N^4}$,  verified numerically in section V.

\section{Calculation of variance of von Neumann entropy from eq.(\ref{pv1})} \label{varr1}

With variance  defined as $\langle \Delta R_1^2 \rangle \equiv \langle R_1^2 \rangle - \langle R_1 \rangle^2$ with  $\langle R_1^n(Y) \rangle =\int R_1^n \,f_v(R_1, Y)\,dR_1$ with $\int f_v \, dR_1=1$,  here again we have
\begin{eqnarray}
 \frac{\partial \langle \Delta R_1^2 \rangle}{\partial Y} &=& \frac{\partial \langle R_1^2 \rangle}{\partial Y} - 2 \langle R_1 \rangle \frac{\partial \langle R_1 \rangle}{\partial Y},
 \label{pvarr1}
\end{eqnarray}

Using the definition of $\langle R_1 \rangle$ and $\langle R_1^2 \rangle$ along with  eq.(\ref{pfv}) into above equation, with $t \approx 1 - R_1^2$, $a \approx 2w, \, b \approx \beta N (N + \nu -1) - \beta \frac{N_{\nu}}{2} \langle R_0 \rangle + N + 2\omega$, and repeating the steps in the preceding section we get,
\begin{eqnarray}
    \frac{\partial \langle R_1 \rangle}{\partial Y} &=& (4-a) \langle R_1 \rangle - (4+b) \\
    \frac{\partial \langle R_1^2 \rangle}{\partial Y} &=& (6-2a) \langle R_1^2 \rangle - 2(4+b)\langle R_1 \rangle + 2.
\end{eqnarray}
Substitution of the above equations in eq.(\ref{pvarr1}), and, as $b$ and $a$ are large, we can approximate 

\begin{equation}
    \frac{\partial \langle \Delta R_1^2 \rangle}{\partial Y} = -2a \langle \Delta R_1^2 \rangle + 2,
\end{equation}
or,
\begin{equation}
    \langle \Delta R_1^2 \rangle = \frac{1}{a} - B e^{-2aY}.
\end{equation}
That is, in the stationary limit $\langle \Delta R_1^2 \rangle \sim \frac{1}{N^2}$, which is much larger that the fluctuations in purity, which we verify numerically.

\end{document}